\documentclass[aps,prd,onecolumn,showpacs,showkeys,amsmath,amssymb]{revtex4}
\usepackage{amsfonts}
\usepackage{amsmath}
\usepackage{graphicx}
\usepackage{subfigure}
\usepackage{dcolumn}
\usepackage{bm}
\usepackage{booktabs}
\usepackage[dvips]{color}
\makeatletter
\newcommand{\figcaption}{\def\@captype{figure}\caption}
\newcommand{\tabcaption}{\def\@captype{table}\caption}

\newcommand{\Rmnum}[1]{\expandafter\@slowromancap\romannumeral #1@}
\def\hlinewd#1{%
  \noalign{\ifnum0=`}\fi\hrule \@height #1 \futurelet
   \reserved@a\@xhline}
\makeatother
\def\dab{\int^{\alpha_{max}}_{\alpha_{min}}d\alpha\int^{\beta_{max}}_{\beta_{min}}d\beta}
\def\qq{\langle\bar qq\rangle}
\def\GGa{\langle GG\rangle}
\def\GGb{\langle g_s^2GG\rangle}
\def\qGq{\langle\bar qg_s\sigma Gq\rangle}
\def\efun{e^{-\frac{m_c^2}{\alpha(1-\alpha)M_B^2}}}
\def\f(s){[(\alpha+\beta)m_c^2-\alpha\beta s]}
\def\non{\\ \nonumber}

\begin{document}

\title{The Vector and Axial-Vector Charmonium-like States}

\author{Wei Chen}
\email{boya@pku.edu.cn} \affiliation{Department of Physics
and State Key Laboratory of Nuclear Physics and Technology\\
Peking University, Beijing 100871, China  }
\author{Shi-Lin Zhu}
\email{zhusl@pku.edu.cn} \affiliation{Department of Physics
and State Key Laboratory of Nuclear Physics and Technology\\
and Center of High Energy Physics, Peking University, Beijing
100871, China }

\begin{abstract}
After constructing all the tetraquark interpolating currents with
$J^{PC}=1^{-+}, 1^{--}, 1^{++}$ and $1^{+-}$ in a systematic way,
we investigate the two-point correlation functions to extract the
masses of the charmonium-like states with QCD sum rule. For the
$1^{--}$ $qc\bar q\bar c$ charmonium-like state, $m_X=4.6\sim4.7$
GeV, which implies a possible tetraquark interpretation for the
state $Y(4660)$. The masses for both the $1^{++}$ $qc\bar q\bar c$
and $sc\bar s\bar c$ charmonium-like states are around $4.0\sim
4.2$ GeV, which are slightly above the mass of $X(3872)$.
For the $1^{-+}$ and $1^{+-}$ $qc\bar q\bar c$
charmonium-like states, the extracted masses are around $4.5\sim
4.7$ GeV and $4.0\sim 4.2$ GeV respectively. As a byproduct, the
bottomonium-like states are also studied. We also discuss the
possible decay modes and experimental search of the
charmonium-like states.
\end{abstract}

\keywords{Charmonium-like states, QCD sum rule}

\pacs{12.38.Lg, 11.40.-q, 12.39.Mk}

\maketitle

\section{Introduction}\label{sec:Introduction}

In the past several years, many unexpected charmonium-like states
have been discovered at B-factories, some of which lie above the
open charm threshold and decay into final states that contain a
$c\bar c$ pair. Some of them do not fit in the conventional quark
model easily and are considered as the candidates of the exotic
states beyond the quark model, such as the molecular states,
tetraquark states, the charmonium hybrid mesons, baryonium states
and so on. For experimental reviews of these new states, one can
consult Refs. ~\cite{2006-Swanson-p243-305, 2008-Zhu-p283-322,
2009-Bracko-p-, 2009-Yuan-p-, 2007-Rosner-p12002-12002}.

The underlying structure of these new states inspired the
extensive study of the hadron spectroscopy. $X(3872)$ is the best
studied charmonium-like state since its discovery by the Belle
Collaboration~\cite{2003-Choi-p262001-262001}. Although the
analysis of angular distributions favors the assignment
$J^{PC}=1^{++}$~\cite{2005-Abe-p-, 2007-Abulencia-p132002-132002},
the $2^{-+}$ possibility is not ruled
out~\cite{2010-AmoSanchez-p11101-11101}. The mass and decay mode
of $X(3872)$ are very different from that of the $2^3P_1$ $c\bar
c$ state. Up to now, the possible interpretations of $X(3872)$
include the molecular state~\cite{2009-Liu-p411-428,
2008-Liu-p63-73, 2004-Swanson-p197-202, 2004-Swanson-p189-195,
2004-Close-p119-123, 2008-Thomas-p34007-34007,
2009-Fernandez-Carames-p222001-222001}, tetraquark
state~\cite{2007-Matheus-p14005-14005, 2007-Maiani-p182003-182003,
2006-Ebert-p214-219}, cusp~\cite{2004-Bugg-p8-14} and hybrid
charmonium~\cite{2003-Close-p210-216}. In
Ref.~\cite{2007-Matheus-p14005-14005}, the authors studied the
$J^{PC}=1^{++}$ state using a tetraquark current in the framework
of the QCD sum rule approach.

The initial state radiation (ISR) process played an important role
in the search of the $1^{--}$ charmonium-like states at
B-factories. BaBar Collaboration first observed $Y(4260)$ in the
$e^+e^-\rightarrow\gamma_{ISR}J/\psi\pi^+\pi^-$
process~\cite{2005-Aubert-p142001-142001}, which was confirmed by
Belle Collaboration~\cite{2007-Yuan-p182004-182004}. Then, BaBar
studied the $\gamma_{ISR}\psi(2S)\pi^+\pi^-$ channel and observed
a broad enhancement around 4.32 GeV. Using the same technique,
Belle observed two distinct resonances $Y(4360)$ and $Y(4660)$ in
the $\psi(2S)\pi^+\pi^-$ mass distribution
~\cite{2007-Wang-p142002-142002}. The masses of these new
charmonium-like states are higher than the open charm threshold.
However, the $Y\rightarrow D^{(\ast)}\bar{D}^{(\ast)}$ decay modes
have not been observed yet~\cite{2007-Abe-p92001-92001}, which are
predicted to be the dominant decay modes of the charmonium above
the open charm threshold in the potential model. In
Refs.~\cite{2010-Wang-p323-332, 2009-Albuquerque-p53-66}, the
authors studied the $1^{--}$ charmonium-like Y mesons using the
QCD sum rule approach. Maiani \textit{et al.} tried to assign
$Y(4260)$ as the $sc\bar s\bar c$ tetraquark in a P-wave
state~\cite{2005-Maiani-p031502-031502}. $Y(4260)$ was also
interpreted as the interesting charmonium hybrid
state~\cite{2005-Zhu-p212-212, 2005-Close-p215-222,
2005-Kou-p164-169}. $Y(4660)$ was considered as a
$\psi(2S)f_0(980)$ bound state~\cite{2008-Guo-p26-29}.

The exotic state with $J^{PC}=1^{-+}$ can not be a $q\bar q$ state
in the simple quark model. Neither can the $1^{-+}$ state be
formed by two gluons due to the Landau-Yang selection
rule~\cite{1948-Landau-p207-207, 1950-Yang-p242-245}. States with
such a quantum number are good candidates of the hybrid meson,
which has been studied with the MIT Bag
model~\cite{1983-Barnes-p241-241, 1983-Chanowitz-p211-211}, the
flux tube model~\cite{1985-Isgur-p2910-2910,
1999-Page-p34016-34016} and the QCD sum rule
formalism~\cite{1984-Govaerts-p1-1, 1987-Latorre-p347-347}. Recently there have been
some efforts on the $1^{-+}$ charmonium-like exotic states. For
example, the structure of $X(4350)$ was studied using a
$D^{\ast}_sD_{s0}^{\ast}$ current with
$J^{PC}=1^{-+}$~\cite{2010-Albuquerque-p-}. Moreover, the newly
observed state $Y(4140)$ was argued as a $1^{-+}$ exotic
charmonium hybrid state~\cite{2009-Mahajan-p228-228}.

In Ref.~\cite{2010-Nielsen-p41-83}, Nielsen \textit{et
al.} reviewed the charmonium-like states from the perspective of
the QCD sum rule approach. They studied new resonances such as
$X(3872), Z^+(4430)$ and $Y(4660)$ etc using one single
molecular-type or tetraquark-type interpolating current. In this
paper, we first construct all the local charmonium-like tetraquark
currents with $J^{PC}=1^{-+}, 1^{--}, 1^{++}$ and $1^{+-}$ in a
systematic way. With these independent currents, we study the
two-point correlation functions and extract the masses of the
possible $1^{-+}, 1^{--}, 1^{++}, 1^{+-}$ states. We study both
the $qQ\bar q\bar Q$ and $sQ\bar s\bar Q$ systems where $Q=c, b$.

The paper is organized as follows. In Sec.~\Rmnum{2}, we construct
the tetraquark interpolating currents with $J^{PC}=1^{-+}, 1^{--},
1^{++}$ and $1^{+-}$ using the diquark and antidiquark fields. In
Sec.~\Rmnum{3}, we calculate the correlation functions and
spectral densities of the interpolating currents and collect them
in the Appendix. We perform the numerical analysis and extract the
masses in Sec.~\Rmnum{4} and discuss the possible decay modes
and experimental search of the charmonium-like states in Sec.~\Rmnum{5}.
The last section is a brief summary.
%
%
\section{TETRAQUARK INTERPOLATING CURRENTS}\label{curr}

In this section, we construct the diquark-antidiquark type of
currents using the technique developed in our previous
works~\cite{2008-Chen-p54017-54017, 2009-Jiao-p114034-114034,
2010-Chen-p105018-105018}. Considering the Lorentz structures,
there are five independent diquark fields (or anti-diquark
fields): $q_a^TC\gamma_5c_b$, $q_a^TCc_b$,
$q_a^TC\gamma_{\mu}\gamma_5c_b$, $q_a^TC\gamma_{\mu}c_b$ and
$q_a^TC\sigma_{\mu\nu}c_b$, where $a$, $b$ are color indices and
$q$ denotes an $up$ or $down$ quark. We will also take account of
$q_a^TC\sigma_{\mu\nu}\gamma_5c_b$ although it is equivalent to
$q_a^TC\sigma_{\mu\nu}c_b$. In fact, they have different parities.
Using these diquarks and anti-diquarks as the basis, we can
compose a six-order matrix $O$. The elements of $O$ are the
tetraquark operators without color structures. We show the spins
and parities of the matrix elements with $J\le 1$ in
Table~\ref{table1}.
\begin{center}
\begin{tabular}{ll|cccccc}
\hlinewd{.8pt}
Operators                           &                                   &   $\bar q_a\gamma_5C\bar c_b^T$
                                    & $\bar q_aC\bar c_b^T$             &   $\bar q_a\gamma_{\mu}\gamma_5C\bar c_b^T$
                                    & $\bar q_a\gamma_{\mu}C\bar c_b^T$ &   $\bar q_a\sigma_{\mu\nu}C\bar c_b^T$
                                    & $\bar q_a\sigma_{\mu\nu}\gamma_5C\bar c_b^T$ \\
                                    & $J^P$ &   $0^+$     &    $0^-$    &    $1^-$    &    $1^+$    &    $1^-$    &    $1^+$ \\
\hline
$q_a^TC\gamma_5c_b$                 & $0^+$ &    $0^+$    &    $0^-$    &    $1^-$    &    $1^+$    &    $ - $    &    $ - $ \\
$q_a^TCc_b$                         & $0^-$ &    $0^-$    &    $0^+$    &    $1^+$    &    $1^-$    &    $ - $    &    $ - $ \\
$q_a^TC\gamma_{\mu}\gamma_5c_b$     & $1^-$ &    $1^-$    &    $1^+$    &    $0^+$    &    $0^-$    &    $1^+$    &    $1^-$ \\
$q_a^TC\gamma_{\mu}c_b$             & $1^+$ &    $1^+$    &    $1^-$    &    $0^-$    &    $0^+$    &    $1^-$    &    $1^+$ \\
$q_a^TC\sigma_{\mu\nu}c_b$          & $1^-$ &    $ - $    &    $ - $    &    $1^+$    &    $1^-$    &    $0^+$    &    $0^-$ \\
$q_a^TC\sigma_{\mu\nu}\gamma_5c_b$  & $1^+$ &    $ - $    &    $ - $    &    $1^-$    &    $1^+$    &    $ - $    &    $ - $ \\
\hlinewd{.8pt}
\end{tabular}
\tabcaption{The spins and parities of the elements of the matrix
$O$.} \label{table1}
\end{center}

We do not use $O_{66}$ and $O_{65}$ in the construction of the
currents since $O_{66}$ is equivalent to $O_{55}$ and $O_{65}$
equivalent to $O_{56}$. One notes that under the
charge-conjugation transformation:
\begin{eqnarray}
\mathbb{C}O\mathbb{C}^{-1}=O^T,
\end{eqnarray}
where $O^T$ is the transpose of $O$. With this relation, we can define the
symmetric matrix $S$ and antisymmetric matrix $A$:
\begin{eqnarray}
S=O+O^T, A=O-O^T
\end{eqnarray}
The elements of these two matrices are the tetraquark operators
that have definite C-parity: they have even and odd C-parities for
the elements of $S$ and $A$, respectively. $A_{ii}=0$ indicates
that the $J^{PC}=0^{+-}$ tetraquark currents without derivatives
do not exist, which has been proven in
Ref.~\cite{2009-Jiao-p114034-114034}.

The diquark and antidiquark should have the same color symmetries
to compose a color singlet tetraquark current. So the color
structure of the tetraquark is either $\mathbf 6 \otimes \mathbf {
\bar 6}$ or $\mathbf { \bar 3 }\otimes \mathbf 3$, which is
denoted by $\mathbf  6 $ and $\mathbf 3 $ respectively. Finally,
we can obtain the tetraquark interpolating currents with
$J^{PC}=1^{-+}, 1^{--}, 1^{++}$ and $1^{+-}$ from the matrices $S$
and $A$:
\begin{itemize}
\item The interpolating currents with $J^{PC}=1^{-+}$ and $1^{--}$
are:
\begin{eqnarray}
\nonumber J_{1\mu}&=&S_{13}^6(A_{13}^6)
=q_{a}^TC\gamma_5c_{b}(\bar{q}_{a}\gamma_{\mu}\gamma_5C\bar{c}^T_{b}+\bar{q}_{b}\gamma_{\mu}\gamma_5C\bar{c}^T_{a})
\pm
q_{a}^TC\gamma_{\mu}\gamma_5c_{b}(\bar{q}_{a}\gamma_5C\bar{c}^T_{b}+\bar{q}_{b}\gamma_5C\bar{c}^T_{a})\,
, \non J_{2\mu}&=&S_{45}^3(A_{45}^3)
=q_{a}^TC\gamma^{\nu}c_{b}(\bar{q}_{a}\sigma_{\mu\nu}C\bar{c}^T_{b}-\bar{q}_{b}\sigma_{\mu\nu}C\bar{c}^T_{a})
\pm
q_{a}^TC\sigma_{\mu\nu}c_{b}(\bar{q}_{a}\gamma^{\nu}C\bar{c}^T_{b}-\bar{q}_{b}\gamma^{\nu}C\bar{c}^T_{a})\,
, \non J_{3\mu}&=&S_{13}^3(A_{13}^3)
=q_{a}^TC\gamma_5c_{b}(\bar{q}_{a}\gamma_{\mu}\gamma_5C\bar{c}^T_{b}-\bar{q}_{b}\gamma_{\mu}\gamma_5C\bar{c}^T_{a})
\pm
q_{a}^TC\gamma_{\mu}\gamma_5c_{b}(\bar{q}_{a}\gamma_5C\bar{c}^T_{b}-\bar{q}_{b}\gamma_5C\bar{c}^T_{a})\,
, \non J_{4\mu}&=&S_{45}^6(A_{45}^6)
=q_{a}^TC\gamma^{\nu}c_{b}(\bar{q}_{a}\sigma_{\mu\nu}C\bar{c}^T_{b}+\bar{q}_{b}\sigma_{\mu\nu}C\bar{c}^T_{a})
\pm
q_{a}^TC\sigma_{\mu\nu}c_{b}(\bar{q}_{a}\gamma^{\nu}C\bar{c}^T_{b}+\bar{q}_{b}\gamma^{\nu}C\bar{c}^T_{a})\,
,
\\ \label{currents1}
J_{5\mu}&=&S_{24}^6(A_{24}^6)
=q_{a}^TCc_{b}(\bar{q}_{a}\gamma_{\mu}C\bar{c}^T_{b}+\bar{q}_{b}\gamma_{\mu}C\bar{c}^T_{a})
\pm
q_{a}^TC\gamma_{\mu}c_{b}(\bar{q}_{a}C\bar{c}^T_{b}+\bar{q}_{b}C\bar{c}^T_{a})\,
, \non J_{6\mu}&=&S_{36}^6(A_{36}^6)
=q_{a}^TC\gamma^{\nu}\gamma_5c_{b}(\bar{q}_{a}\sigma_{\mu\nu}\gamma_5C\bar{c}^T_{b}+\bar{q}_{b}\sigma_{\mu\nu}\gamma_5C\bar{c}^T_{a})
\pm
q_{a}^TC\sigma_{\mu\nu}\gamma_5c_{b}(\bar{q}_{a}\gamma^{\nu}\gamma_5C\bar{c}^T_{b}+\bar{q}_{b}\gamma^{\nu}
\gamma_5C\bar{c}^T_{a})\, , \non J_{7\mu}&=&S_{24}^3(A_{24}^3)
=q_{a}^TCc_{b}(\bar{q}_{a}\gamma_{\mu}C\bar{c}^T_{b}-\bar{q}_{b}\gamma_{\mu}C\bar{c}^T_{a})
\pm
q_{a}^TC\gamma_{\mu}c_{b}(\bar{q}_{a}C\bar{c}^T_{b}-\bar{q}_{b}C\bar{c}^T_{a})\,
, \non J_{8\mu}&=&S_{36}^3(A_{36}^3)
=q_{a}^TC\gamma^{\nu}\gamma_5c_{b}(\bar{q}_{a}\sigma_{\mu\nu}\gamma_5C\bar{c}^T_{b}-\bar{q}_{b}\sigma_{\mu\nu}\gamma_5C\bar{c}^T_{a})
\pm
q_{a}^TC\sigma_{\mu\nu}\gamma_5c_{b}(\bar{q}_{a}\gamma^{\nu}\gamma_5C\bar{c}^T_{b}-\bar{q}_{b}\gamma^{\nu}
\gamma_5C\bar{c}^T_{a})\, .
\end{eqnarray}
where ``$S$'' and ``$+$'' correspond to $J^{PC}=1^{-+}$, ``$A$''
and ``$-$'' correspond to $J^{PC}=1^{--}$.

\item The interpolating currents with $J^{PC}=1^{++}$ and $1^{+-}$
are:
\begin{eqnarray}
\nonumber J_{1\mu}&=&S_{23}^6(A_{23}^6)
=q^T_aCc_b(\bar{q}_a\gamma_{\mu}\gamma_5C\bar{c}^T_b+\bar{q}_b\gamma_{\mu}\gamma_5C\bar{c}^T_a)
\pm
q^T_aC\gamma_{\mu}\gamma_5c_b(\bar{q}_aC\bar{c}^T_b+\bar{q}_bC\bar{c}^T_a)\,
, \non J_{2\mu}&=&S_{23}^3(A_{23}^3)
=q^T_aCc_b(\bar{q}_a\gamma_{\mu}\gamma_5C\bar{c}^T_b-\bar{q}_b\gamma_{\mu}\gamma_5C\bar{c}^T_a)
\pm
q^T_aC\gamma_{\mu}\gamma_5c_b(\bar{q}_aC\bar{c}^T_b-\bar{q}_bC\bar{c}^T_a)\,
, \non J_{3\mu}&=&S_{14}^6(A_{14}^6)
=q^T_aC\gamma_5c_b(\bar{q}_a\gamma_{\mu}C\bar{c}^T_b+\bar{q}_b\gamma_{\mu}C\bar{c}^T_a)
\pm
q^T_aC\gamma_{\mu}c_b(\bar{q}_a\gamma_5C\bar{c}^T_b+\bar{q}_b\gamma_5C\bar{c}^T_a)\,
,
\\ \label{currents2}
J_{4\mu}&=&S_{14}^3(A_{14}^3)
=q^T_aC\gamma_5c_b(\bar{q}_a\gamma_{\mu}C\bar{c}^T_b-\bar{q}_b\gamma_{\mu}C\bar{c}^T_a)
\pm
q^T_aC\gamma_{\mu}c_b(\bar{q}_a\gamma_5C\bar{c}^T_b-\bar{q}_b\gamma_5C\bar{c}^T_a)\,
, \non J_{5\mu}&=&S_{46}^6(A_{46}^6)
=q^T_aC\gamma^{\nu}c_b(\bar{q}_a\sigma_{\mu\nu}\gamma_5C\bar{c}^T_b+\bar{q}_b\sigma_{\mu\nu}\gamma_5C\bar{c}^T_a)
\pm
q^T_aC\sigma_{\mu\nu}\gamma_5c_b(\bar{q}_a\gamma^{\nu}C\bar{c}^T_b+\bar{q}_b\gamma^{\nu}C\bar{c}^T_a)\,
, \non J_{6\mu}&=&S_{46}^3(A_{46}^3)
=q^T_aC\gamma^{\nu}c_b(\bar{q}_a\sigma_{\mu\nu}\gamma_5C\bar{c}^T_b-\bar{q}_b\sigma_{\mu\nu}\gamma_5C\bar{c}^T_a)
\pm
q^T_aC\sigma_{\mu\nu}\gamma_5c_b(\bar{q}_a\gamma^{\nu}C\bar{c}^T_b-\bar{q}_b\gamma^{\nu}C\bar{c}^T_a)\,
, \non J_{7\mu}&=&S_{35}^6(A_{35}^6)
=q^T_aC\gamma^{\nu}\gamma_5c_b(\bar{q}_a\sigma_{\mu\nu}C\bar{c}^T_b+\bar{q}_b\sigma_{\mu\nu}C\bar{c}^T_a)
\pm
q^T_aC\sigma_{\mu\nu}c_b(\bar{q}_a\gamma^{\nu}\gamma_5C\bar{c}^T_b+\bar{q}_b\gamma^{\nu}\gamma_5C\bar{c}^T_a)\,
, \non J_{8\mu}&=&S_{35}^3(A_{35}^3)
=q^T_aC\gamma^{\nu}\gamma_5c_b(\bar{q}_a\sigma_{\mu\nu}C\bar{c}^T_b-\bar{q}_b\sigma_{\mu\nu}C\bar{c}^T_a)
\pm
q^T_aC\sigma_{\mu\nu}c_b(\bar{q}_a\gamma^{\nu}\gamma_5C\bar{c}^T_b-\bar{q}_b\gamma^{\nu}\gamma_5C\bar{c}^T_a)\,
.
\end{eqnarray}
where ``$S$'' and ``$+$'' correspond to $J^{PC}=1^{++}$, ``$A$''
and ``$-$'' correspond to $J^{PC}=1^{+-}$.
\end{itemize}
In order to have definite isospin and $G$-parity, all the currents
in Eqs.~(\ref{currents1})-(\ref{currents2}) should contain
$(uc\bar u \bar c + dc\bar d \bar c)$. However, we do not
differentiate the $up$ and $down$ quarks in our analysis due to
the isospin symmetry and denote them by $q$.

%
%
\section{SPECTRAL DENSITY}\label{sec:QSR}
In the past several decades, QCD sum rule has been widely used to
study the hadron structures and proven to be a very powerful
non-perturbative method~\cite{1979-Shifman-p385-447,
1985-Reinders-p1-1, 2000-Colangelo-p-}. We consider the two-point
correlation function:
\begin{eqnarray}
\nonumber \Pi_{\mu\nu}(q^{2})&=& i\int
d^4xe^{iqx}\langle0|T[J_{\mu}(x)J_{\nu}^{\dag}(0)]|0\rangle
\\
&=&-\Pi_1(q^2)(g_{\mu\nu}-\frac{q_{\mu}q_{\nu}}{q^2})+\Pi_0(q^2)\frac{q_{\mu}q_{\nu}}{q^2},\label{equ:Pi}
\end{eqnarray}
There are two independent parts of $\Pi_{\mu\nu}$ with different
Lorentz structures because $J_{\mu}$ is not a conserved current.
$\Pi_1(q^2)$ is related to the vector meson while $\Pi_0(q^2)$ is
the scalar current polarization function.

At the hadron level, the correlation function is expressed by the
dispersion relation with a spectral function:
\begin{eqnarray}
\Pi_1(q^2)=\int_{4m_c^2}^{\infty}\frac{\rho(s)}{s-q^2-i\epsilon},
\label{Phenpi}
\end{eqnarray}
where the lower limit of integration is the square of the sum of
the mass of all current quarks (omitting the light quark mass).
The spectral function is defined as:
\begin{eqnarray}
\nonumber
\rho(s)&\equiv&\sum_n\delta(s-m_n^2)\langle0|\eta|n\rangle\langle n|\eta^+|0\rangle\\
&=&f_X^2\delta(s-m_X^2)+ \mbox{continuum},   \label{Phenrho}
\end{eqnarray}
where the usual pole plus continuum parametrization of the
hadronic spectral density is adopted.

On the other hand, the correlation function can also be calculated
at the quark-gluon level via the operator product expansion(OPE)
method. Using the same technique as in
Refs.~\cite{2010-Albuquerque-p-, 2008-Lee-p28-32,
2007-Matheus-p14005-14005, 2010-Chen-p105018-105018}, we evaluate
the Wilson coefficient up to dimension eight while the coordinate
space expression for the light quark propagator and the momentum
space expression for the charm quark propagator are adopted:
\begin{eqnarray}
\nonumber iS^{ab}_q(x) &=&  \frac{i\delta^{ab}}{2\pi^2x^4}\hat{x}
+\frac{i}{32\pi^2}\frac{\lambda^n_{ab}}{2}g_sG_{\mu\nu}^n\frac{1}
{x^2}(\sigma^{\mu\nu}\hat{x}+\hat{x}\sigma^{\mu\nu})-\frac{\delta^{ab}}{12}\langle\bar{q}q\rangle+
\frac{\delta^{ab}x^2}{192}\qGq,
\\
iS^{ab}_c(p) &=&
\frac{i\delta^{ab}}{\hat{p}-m_c}+\frac{i}{4}g_s\frac{\lambda^n_{ab}}{2}G_{\mu\nu}^n
\frac{\sigma^{\mu\nu}(\hat{p}+m_c)+(\hat{p}+m_c)\sigma^{\mu\nu}}
{(p^2-m_c^2)^2}+\frac{i\delta^{ab}}{12}\GGb
m_c\frac{p^2+m_c\hat{p}}{(p^2-m_c^2)^4}.
\end{eqnarray}
where $\hat{x}\equiv\gamma_{\mu}x^{\mu}$,
$\hat{p}\equiv\gamma_{\mu}p^{\mu}$, $\qGq= \langle g_s
\bar{q}\sigma^{\mu\nu} G_{\mu\nu}q\rangle$, $\GGb=\langle
g_s^2G_{\mu\nu}G^{\mu\nu}\rangle$, $a$ and $b$ are color indices.
The dimensional regularization is used throughout our calculation.
The spectral density is then obtained with:
$\rho(s)=\frac{1}{\pi}$Im$\Pi(q^2)$.

One of the important assumption in the QCD sum rule approach is
the quark-hadron duality, which ensures the equivalence of the
correlation functions obtained at the hadron level and the
quark-gluon level. To improve the convergence of the OPE series,
the Borel transformation is performed to the correlation functions
at both levels. Considering the spectral function defined in
Eq.~(\ref{Phenrho}), we arrive at:
\begin{eqnarray}
f_X^2e^{-m_X^2/M_B^2}=\int_{4m_c^2}^{s_0}dse^{-s/M_B^2}\rho(s),
\label{sumrule}
\end{eqnarray}
where $s_0$ is the threshold parameter. Then we can extract the
mass $m_X$:
\begin{eqnarray}
m_X^2=\frac{\int_{4m_c^2}^{s_0}dse^{-s/M_B^2}s\rho(s)}{\int_{4m_c^2}^{s_0}dse^{-s/M_B^2}\rho(s)}.
\label{mass}
\end{eqnarray}

The spectral extracted densities of all the tetraquark currents in
Eq.~(\ref{currents1}) are listed in the Appendix. For each
quantum number, we just list the expressions of four currents.
Others could be obtained conveniently by simple replacement
$m_c\rightarrow -m_c$. We collect the terms that are proportional
to the light quark mass $m_q$ into the expressions of the spectral
densities. These terms give small contributions to the correlation
functions involving the $qc\bar q\bar c$ system and hence can be
ignored. In the case of the $sc\bar s\bar c$ system, however, they
give important corrections because the strange quark mass $m_s$ is
much larger than $m_u$ and $m_d$. Both the quark condensate $\qq$
and quark gluon mixed condensate $\qGq$ appear with both $m_q$ and
$m_c$, which is very different from the case in the pseudoscalar
channel~\cite{2010-Chen-p105018-105018}, where only the $m_q$
related terms contribute. We neglect the three gluon condensate
$g_s^3\langle fGGG\rangle$ because they are strongly suppressed
and negligible~\cite{2010-Chen-p105018-105018}.

\section{QCD Sum Rule Analysis}\label{sec:analysis}

We perform the QCD sum rule analysis using the following parameter
values of the quark masses and various
condensates~\cite{1979-Shifman-p385-447, 2010-Nakamura-p75021-75021,
2001-Eidemuller-p203-210, 1999-Jamin-p300-303,
2002-Jamin-p237-243}:
\begin{eqnarray}
\nonumber &&m_u(2\text{ GeV})=(2.9\pm0.6)\text{ MeV} \, , \non
&&m_d(2\text{ GeV})=(5.2\pm0.9)\text{ MeV} \, , \non &&m_q(2\text{
GeV})=(4.0\pm0.7)\text{ MeV} \, , \non &&m_s(2\text{
GeV})=(101^{+29}_{-21})\text{ MeV} \, , \non
&&m_c(m_c)=(1.23\pm0.09)\text{ GeV} \, , \non
&&m_b(m_b)=(4.20\pm0.07)\text{ GeV} \, , \non
&&\qq=-(0.23\pm0.03)^3\text{ GeV}^3 \, , \non &&\qGq=-M_0^2\qq\, ,
\non &&M_0^2=(0.8\pm0.2)\text{ GeV}^2 \, ,
\\
&&\langle\bar ss\rangle/\qq=0.8\pm0.2 \, , \non &&\GGb=0.88\text{
GeV}^4 \, . \label{parameters}
\end{eqnarray}
where the $up, down$ and $strange$ quark masses are the current
quark masses in a mass-independent subtraction scheme such as
$\overline{MS}$ at a scale $\mu=2$ GeV. The $charm$ and $bottom$
quark masses are the running masses in the $\overline{MS}$ scheme.

\subsection{The tetraquark systems with $J^{PC}=1^{-+}$ and $1^{--}$}

The stability of QCD sum rule requires a suitable working region
of the threshold value $s_0$ and the Borel mass $M_B$. In our
analysis, we choose the value of $s_0$ around which the variation
of the extracted mass $m_X$ with $M_B^2$ is minimum. The working
region of the Borel mass is determined by the convergence of the
OPE series and the pole contribution. The requirement of the
convergence of the OPE series leads to the lower bound $M^2_{min}$
of the Borel parameter while the constraint of the pole
contribution yields the upper bound of $M_B^2$.

We first study the interpolating currents with $J^{PC}=1^{-+}$ and
$1^{--}$ in the $qc\bar q\bar c$ systems. For these currents, the
contribution of the four quark condensate $\qq^2$ is negative and
its absolute value is bigger than other condensates in the region
of $M_B^2<3.1$ GeV$^2$. It is the dominant power contribution to
the correlation function in this region. In fact, the quark
condensate $\qq$ in the spectral density of the $1^{--}$ currents
is proportional to the light quark mass $m_q$ and vanishes if we
take $m_q=0$. For the $1^{-+}$ currents, we require that the
perturbative term be larger than five times of the four quark
condensate to ensure the convergence of the OPE series, which
results in the lower bound of the Borel parameter, $M^2_{min}\sim
2.9$ GeV$^2$. The constraint for the $1^{--}$ currents is stricter due
to the vanishing $\qq$ condensate. The pole contribution (PC) is
defined as:
\begin{eqnarray}
PC=\frac{\int_{4m_c^2}^{s_0}dse^{-s/M_B^2}\rho(s)}{\int_{4m_c^2}^{\infty}dse^{-s/M_B^2}\rho(s)},
\label{pc}
\end{eqnarray}
It is clear that the PC depends on the threshold value $s_0$. As
mentioned above, $s_0$ is chosen to ensure the minimum variation
of $m_X$ with $M_B^2$. For example, we choose $s_0\sim26$ GeV$^2$
for $J_{6\mu}$ with $J^{PC}=1^{-+}$ in Fig.~\ref{fig1}a. By
requiring the pole contribution be larger than $40\%$, we obtain
the upper limit $M^2_{max}$ of the Borel parameter.
\begin{center}
\begin{tabular}{lr}
\scalebox{0.67}{\includegraphics{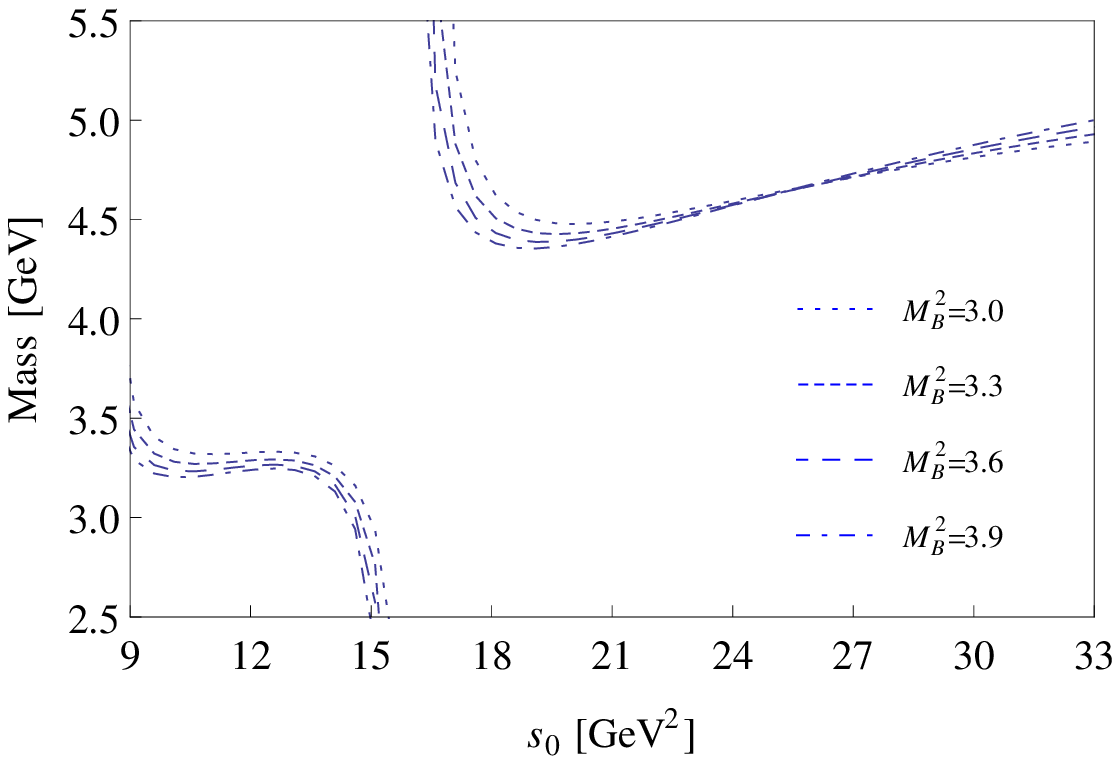}}&
\scalebox{0.67}{\includegraphics{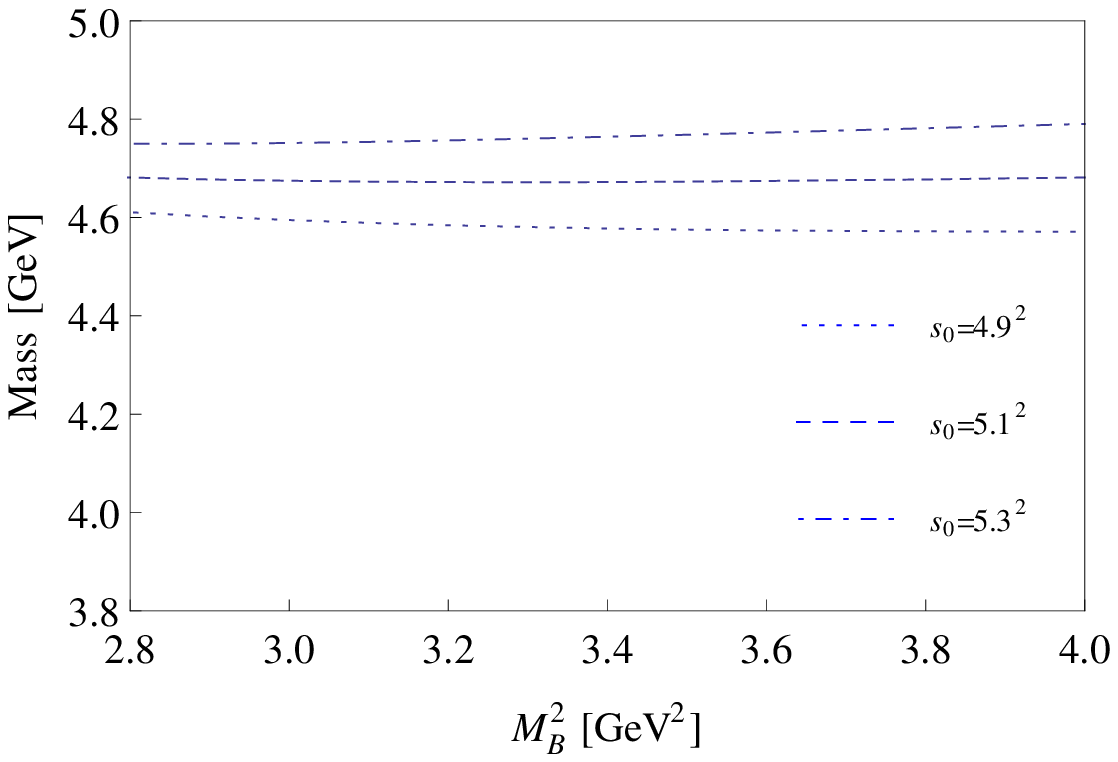}}
\end{tabular}
\centerline{\hspace{0.00in} {(a)} \hspace{3in}{ (b)}}
\figcaption{The variation of $m_X$ with $s_0$(a) and $M^2_B$(b)
corresponding to the current $J_{6\mu}$ for the $1^{-+}$ $qc\bar
q\bar c$ system.} \label{fig1}
\end{center}
\begin{center}
\begin{tabular}{lr}
\scalebox{0.67}{\includegraphics{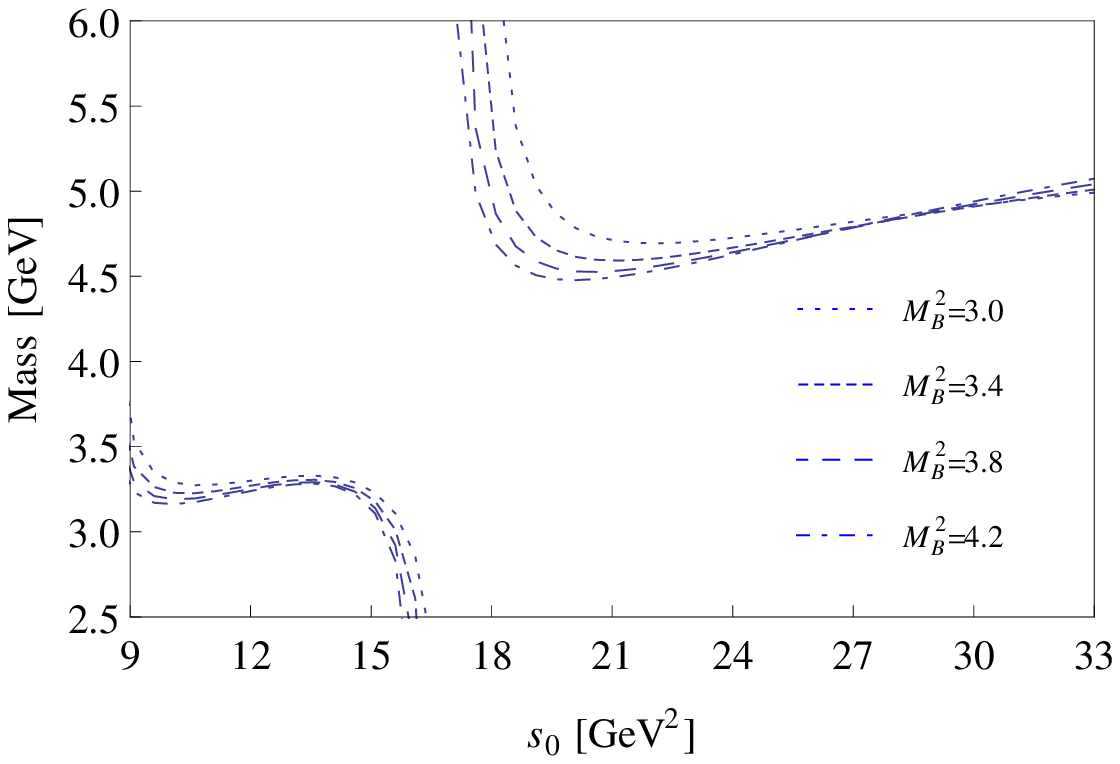}}&
\scalebox{0.67}{\includegraphics{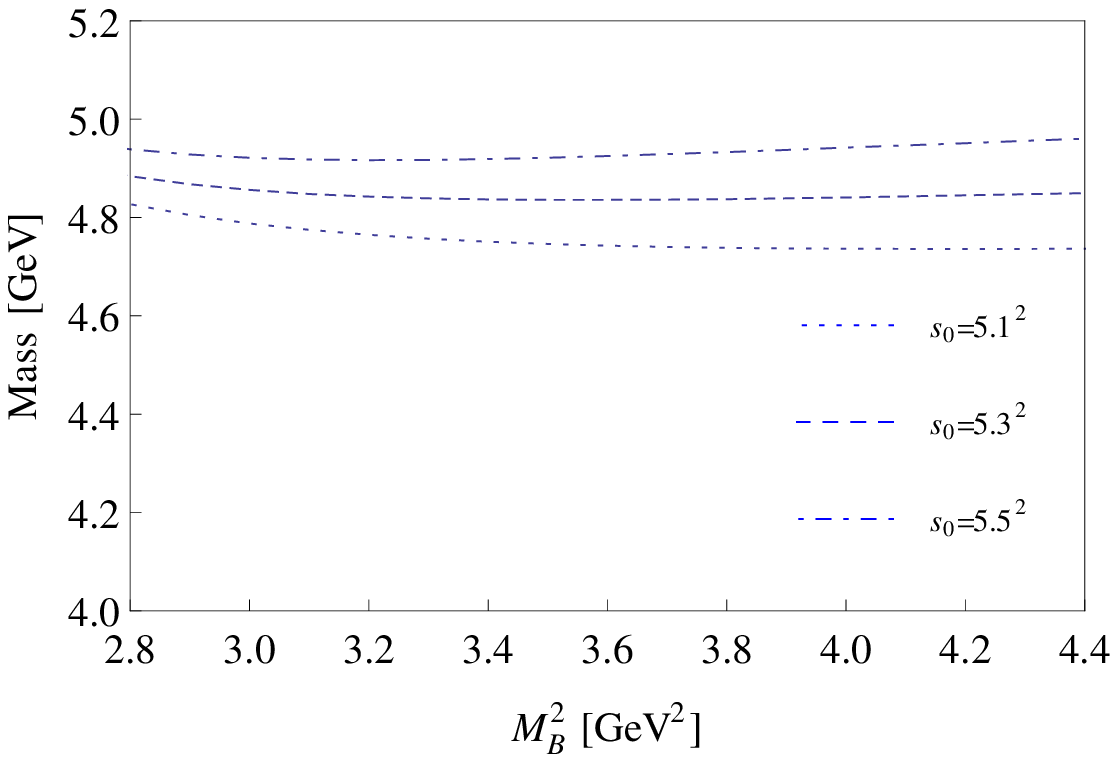}}
\end{tabular}
\centerline{\hspace{0.05in} {(a)} \hspace{3in}{ (b)}}
\figcaption{The variation of $m_X^s$ with $s_0$(a) and $M^2_B$(b)
corresponding to the current $J_{6\mu}$ for the $1^{-+}$ $sc\bar
s\bar c$ system.} \label{fig2}
\end{center}

After performing the QCD sum rule analysis in the working range of
the parameters obtained above, only the currents $J_{6\mu},
J_{7\mu}$ and $J_{8\mu}$ with $J^{PC}=1^{-+}$ have the stable mass
sum rules. In Fig.~\ref{fig1}, we show the variation of $m_X$ with
the threshold value $s_0$ and Borel parameter $M^2_B$ for the
current $J_{6\mu}$ in $qc\bar q\bar c$ system. The plateau in the
region of $10\sim14$ GeV$^2$ in Fig.~\ref{fig1}a is just an
unphysical artifact because both the numerator and denominator in
Eq.~(\ref{mass}) are negative within this region. The situation
also occurs in the pseudoscalar channel in
Ref.~\cite{2010-Chen-p105018-105018}. The variation of $m_X$ with
the Borel parameter $M_B^2$ is weak around the region
$s_0\sim5.1^2$ GeV$^2$, as shown in Fig.~\ref{fig1}b. For
$J_{1\mu}, J_{2\mu}, J_{3\mu}, J_{4\mu}$ and $J_{5\mu}$, the
stability is so bad that the extracted mass $m_X$ grows
monotonically with the threshold value $s_0$ and the Borel
parameter $M_B$. These currents may couple to the $1^{-+}$ states
very weakly and the continuum contribution may be quite large,
leading to the above unstable mass sum rules.

We show the Borel window, the threshold value, the extracted mass
and the pole contribution corresponding to the tetraquark currents
$J_{6\mu}\sim J_{8\mu}$ with $J^{PC}=1^{-+}$ in the $qc\bar q\bar c$
system in Table~\ref{table2}. The results of the $1^{--}$ system
are listed in Table~\ref{table3}. We present the numerical results
for the currents which lead to the stable mass sum rules in the
working range of the Borel parameter. Only the errors
from the uncertainty of the threshold values and variation of the
Borel parameter are taken into account. Other possible error
sources include the truncation of the OPE series and the
uncertainty of the quark masses, condensate values and so on.
%
\begin{center}
\begin{tabular}{cccccc}
\hlinewd{.8pt}
                   & Currents & $s_0(\mbox{GeV}^2)$&$[M^2_{\mbox{min}}$,$M^2_{\mbox{max}}](\mbox{GeV}^2)$&$m_X$\mbox{(GeV)}&PC(\%)\\
\hline
                        & $J_{6\mu}$      &  $5.1^2$             & $2.9\sim3.9$              & $4.67\pm0.10$    & 50.2  \\
$qc\bar q\bar c$ system & $J_{7\mu}$      &  $5.2^2$             & $2.9\sim4.2$              & $4.77\pm0.10$    & 47.4  \\
                        & $J_{8\mu}$      &  $4.9^2$             & $2.9\sim3.4$              & $4.53\pm0.10$    & 46.3
\vspace{5pt}\\
                        & $J_{1\mu}$      &  $5.0^2$             & $2.9\sim3.4$              & $4.67\pm0.10$    & 44.3  \\
                        & $J_{2\mu}$      &  $5.0^2$             & $2.9\sim3.4$              & $4.65\pm0.09$    & 45.6  \\
                        & $J_{3\mu}$      &  $4.9^2$             & $2.9\sim3.3$              & $4.54\pm0.10$    & 44.4  \\
                        & $J_{4\mu}$      &  $5.1^2$             & $2.9\sim3.7$              & $4.72\pm0.09$    & 44.8  \\
$sc\bar s\bar c$ system & $J_{5\mu}$      &  $5.0^2$             & $2.9\sim3.6$              & $4.62\pm0.10$    & 42.8  \\
                        & $J_{6\mu}$      &  $5.3^2$             & $2.9\sim4.3$              & $4.84\pm0.10$    & 47.3  \\
                        & $J_{7\mu}$      &  $5.3^2$             & $2.9\sim4.3$              & $4.87\pm0.10$    & 46.2  \\
                        & $J_{8\mu}$      &  $5.2^2$             & $2.9\sim4.1$              & $4.77\pm0.10$    & 44.1
\vspace{5pt}\\
                        & $J_{6\mu}$      &  $11.0^2$            & $7.2\sim8.6$              & $10.53\pm0.11$   & 44.2  \\
$qb\bar q\bar b$ system & $J_{7\mu}$      &  $11.0^2$            & $7.2\sim8.6$              & $10.53\pm0.10$   & 44.1  \\
                        & $J_{8\mu}$      &  $11.0^2$            & $7.2\sim8.6$              & $10.49\pm0.11$   & 44.7
\vspace{5pt}\\
                        & $J_{4\mu}$      &  $11.0^2$            & $7.2\sim8.1$              & $10.62\pm0.10$   & 41.2  \\
                        & $J_{5\mu}$      &  $11.0^2$            & $7.2\sim8.4$              & $10.56\pm0.10$   & 43.8  \\
$qb\bar q\bar b$ system & $J_{6\mu}$      &  $11.0^2$            & $7.2\sim8.3$              & $10.63\pm0.10$   & 42.4  \\
                        & $J_{7\mu}$      &  $11.0^2$            & $7.2\sim8.3$              & $10.62\pm0.09$   & 42.5  \\
                        & $J_{8\mu}$      &  $11.0^2$            & $7.2\sim8.3$              & $10.59\pm0.10$   & 43.1  \\
\hlinewd{.8pt}
\end{tabular}
\tabcaption{The threshold value, Borel window, mass and pole
contribution corresponding to the currents with $J^{PC}=1^{-+}$ in the
$qc\bar q\bar c$, $sc\bar s\bar c$, $qb\bar q\bar b$ and $sb\bar
s\bar b$ systems.\label{table2}}
\end{center}

The analysis can easily be extended to the $sc\bar s\bar c$
systems. We keep the $m_s$ related terms in the spectral
densities. These terms give important corrections to the OPE
series and are propitious to enhance the stability of the sum
rule. Especially for the currents with $J^{PC}=1^{--}$,
$\langle\bar ss\rangle$ is now proportional to the strange quark
$m_s$ and larger than $\langle\bar ss\rangle^2$ in the Borel
window, which is very different from the $qc\bar q\bar c$ system.
Using the parameters in Eq.~(\ref{parameters}), we also collect
the numerical results for the $1^{--}$ and $1^{-+}$ $sc\bar s\bar
c$ systems in Table~\ref{table2} and Table~\ref{table3}
respectively. Obviously, the $sc\bar s\bar c$ systems have better
stabilities than the $qc\bar q\bar c$ systems. We show the
variation of $m_{X^s}$ with $s_0$ and $M^2_B$ for the current
$J_{6\mu}$ with $J^{PC}=1^{-+}$ in Fig.~\ref{fig2}, which is very
similar to Fig.~\ref{fig1} except the value of $s_0$ around which
the variation of $m_X^s$ with $M_B^2$ is minimum. The extracted
mass of the $sc\bar s\bar c$ state is a little higher than that of
the $qc\bar q\bar c$ state. The mass difference $m_{X^s}-m_X$ for
the same interpolating current is about 0.2 GeV, which is about
$2(m_s-m_q)$ within the errors.

The extracted mass of the $qc\bar q\bar c$ state with
$J^{PC}=1^{--}$ in Table~\ref{table3} is about $4.6\sim4.7$ GeV,
which is consistent with the mass of the meson $Y(4660)$. One may
wonder whether $Y(4660)$ could be a tetraquark state.

Properties of the bottomonium-like analogues tend to be very
similar because of the heavy quark symmetry. Replacing $m_c$ with
$m_b$ in the correlation functions and repeating the same analysis
procedures done above, we collect the relevant results of the $qb\bar
q\bar b$ and $sb\bar s\bar b$ systems in Table~\ref{table2} and
Table~\ref{table3}. One notes that the bottomonium-like systems
are less stable than the corresponding charmonium-like systems.

\begin{center}
\begin{tabular}{cccccc}
\hlinewd{.8pt}
                   & Currents & $s_0(\mbox{GeV}^2)$&$[M^2_{\mbox{min}}$,$M^2_{\mbox{max}}](\mbox{GeV}^2)$&$m_X$\mbox{(GeV)}&PC(\%)\\
\hline
                        & $J_{1\mu}$      &  $5.0^2$         & $2.9\sim3.6$           & $4.64\pm0.09$     & 44.1  \\
$qc\bar q\bar c$ system & $J_{4\mu}$      &  $5.0^2$         & $2.9\sim3.6$           & $4.61\pm0.10$     & 46.4  \\
                        & $J_{7\mu}$      &  $5.2^2$         & $2.9\sim4.1$           & $4.74\pm0.10$     & 47.3
\vspace{5pt} \\
                        & $J_{1\mu}$      &  $5.4^2$         & $2.8\sim4.5$           & $4.92\pm0.10$     & 50.3  \\
                        & $J_{2\mu}$      &  $5.0^2$         & $2.8\sim3.5$           & $4.64\pm0.09$     & 48.6  \\
$sc\bar s\bar c$ system & $J_{3\mu}$      &  $4.9^2$         & $2.8\sim3.4$           & $4.52\pm0.10$     & 45.6  \\
                        & $J_{4\mu}$      &  $5.4^2$         & $2.8\sim4.5$           & $4.88\pm0.10$     & 51.7  \\
                        & $J_{7\mu}$      &  $5.3^2$         & $2.8\sim4.3$           & $4.86\pm0.10$     & 46.0  \\
                        & $J_{8\mu}$      &  $4.8^2$         & $2.8\sim3.1$           & $4.48\pm0.10$     & 43.2
\vspace{5pt} \\
$qb\bar q\bar b$ system & $J_{7\mu}$      &  $11.0^2$        &
$7.2\sim8.5$           & $10.51\pm0.10$    & 45.8
\vspace{5pt} \\
                        & $J_{1\mu}$      &  $11.0^2$        & $7.2\sim8.3$           & $10.60\pm0.10$    & 47.0  \\
                        & $J_{2\mu}$      &  $11.0^2$        & $7.2\sim8.4$           & $10.55\pm0.11$    & 43.6  \\
$sb\bar s\bar b$ system & $J_{3\mu}$      &  $11.0^2$        & $7.2\sim8.4$           & $10.55\pm0.10$    & 43.7  \\
                        & $J_{4\mu}$      &  $11.0^2$        & $7.2\sim8.4$           & $10.53\pm0.11$    & 44.3  \\
                        & $J_{7\mu}$      &  $11.0^2$        & $7.2\sim8.2$           & $10.62\pm0.10$    & 42.0  \\
                        & $J_{8\mu}$      &  $11.0^2$        & $7.2\sim8.4$           & $10.53\pm0.10$    & 44.1  \\
\hlinewd{.8pt}
\end{tabular}
\tabcaption{The threshold value, Borel window, mass and pole
contribution corresponding to the currents with $J^{PC}=1^{--}$ in the
$qc\bar q\bar c$, $sc\bar s\bar c$, $qb\bar q\bar b$ and $sb\bar
s\bar b$ systems.\label{table3}}
\end{center}
\subsection{The tetraquark systems with $J^{PC}=1^{++}$ and $1^{+-}$}

We study the currents with $J^{PC}=1^{++}$ and $1^{+-}$ in this
subsection. The spectral densities of these currents are very
similar to that of the $1^{-+}$ and $1^{--}$ currents, as shown in
the Appendix. The analysis shows that the OPE convergence becomes
worse than that in the vector channel. In this channel, the quark
condensate $\qq$ is larger than any other condensates for all the
currents. The OPE convergence of the currents $J_{5\mu}, J_{6\mu},
J_{7\mu}, J_{8\mu}$ is a little better than that of $J_{1\mu},
J_{2\mu}, J_{3\mu}, J_{4\mu}$.

Only the currents $J_{3\mu}$ and $J_{4\mu}$ in the $1^{++}$
$qc\bar q\bar c$ system display stable mass sum rules. We obtain
the working region of the Borel parameter in $3.0\leq
M_B^2\leq3.4$ GeV$^2$ while taking $s_0=4.6^2$ GeV$^2$ for current
$J_{4\mu}$. The variations of $m_X$ with the threshold value $s_0$
and Borel parameter $M_B^2$ are shown in Fig.~\ref{fig3}, from
which the $M_B^2$ dependence is very weak around the chosen
threshold values. Taking into account only the errors from the
variation of $M_B$ and $s_0$, the extracted mass is $m_X=4.03$
GeV, which is slightly above the mass of $X(3872)$ within the
errors.

The $sc\bar s\bar c$, $qb\bar q\bar b$ and $sb\bar s\bar b$
systems can be studied conveniently by replacement of the
parameters, including the quark masses and the various
condensates. The numerical results are listed in
Table~\ref{table4} for the $1^{++}$ systems and Table~\ref{table5}
for the $1^{+-}$ systems. Now the bottomonium-like systems are
more stable than the corresponding charmonium-like systems.

\begin{center}
\begin{tabular}{lr}
\scalebox{0.67}{\includegraphics{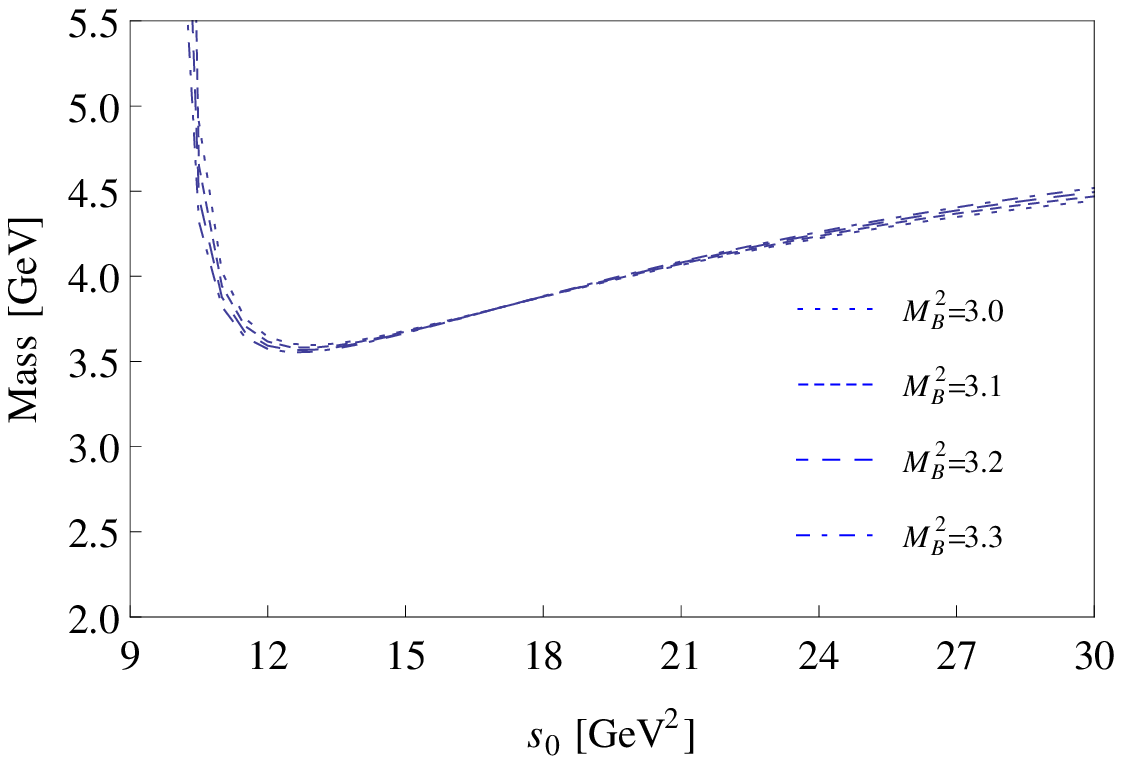}}&
\scalebox{0.67}{\includegraphics{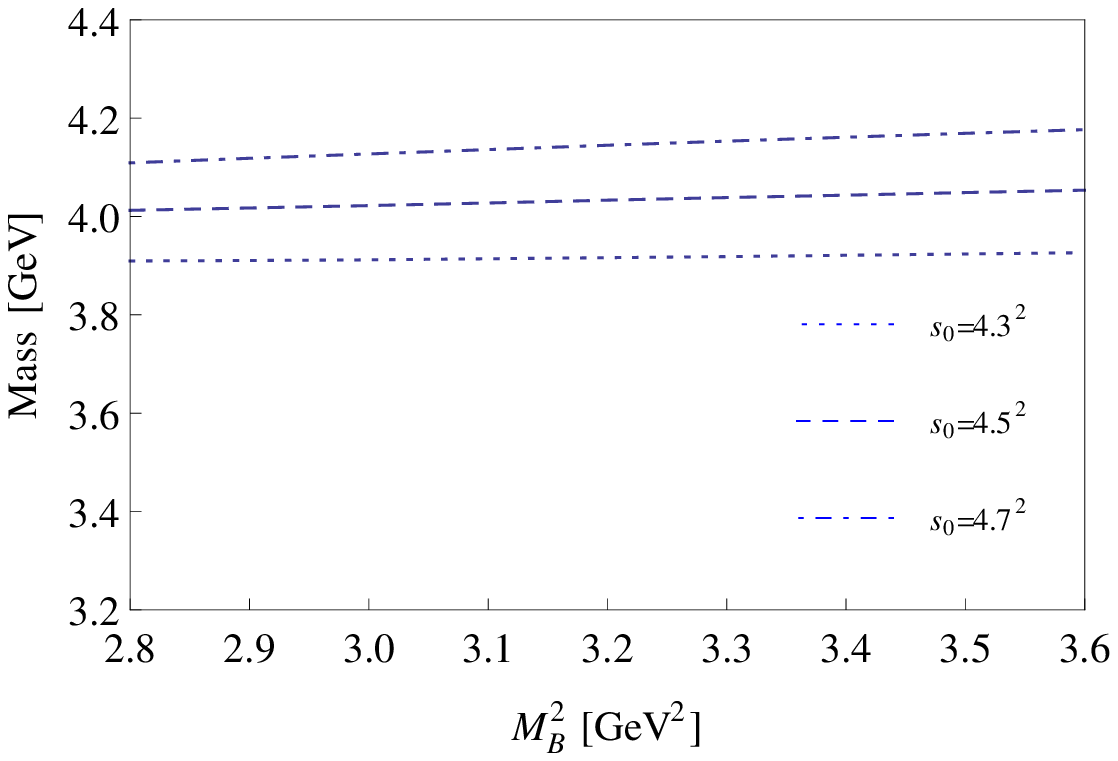}}
\end{tabular}
\centerline{\hspace{0.00in} {(a)} \hspace{3in}{ (b)}}
\figcaption{The variation of $m_X$ with $s_0$(a) and $M^2_B$(b)
corresponding to the current $J_{4\mu}$ for the $1^{++}$ $qc\bar
q\bar c$ system.} \label{fig3}
\end{center}

\begin{center}
\begin{tabular}{cccccc}
\hlinewd{.8pt}
                   & Currents & $s_0(\mbox{GeV}^2)$&$[M^2_{\mbox{min}}$,$M^2_{\mbox{max}}](\mbox{GeV}^2)$&$m_X$\mbox{(GeV)}&PC(\%)\\
\hline
$qc\bar q\bar c$ system & $J_{3\mu}$         &  $4.6^2$         & $3.0\sim3.4$           & $4.19\pm0.10$     & 47.3 \\
                        & $J_{4\mu}$         &  $4.5^2$         & $3.0\sim3.3$           & $4.03\pm0.11$     & 46.8
\vspace{5pt} \\
$sc\bar s\bar c$ system & $J_{3\mu}$         &  $4.6^2$         & $3.0\sim3.4$           & $4.22\pm0.10$     & 45.7  \\
                        & $J_{4\mu}$         &  $4.5^2$         & $3.0\sim3.3$           & $4.07\pm0.10$     & 44.4
\vspace{5pt} \\
                        & $J_{3\mu}$         &  $10.9^2$        & $8.5\sim9.5$           & $10.32\pm0.09$    & 47.0 \\
$qb\bar q\bar b$ system & $J_{4\mu}$         &  $10.8^2$        & $8.5\sim9.2$           & $10.22\pm0.11$    & 44.6 \\
                        & $J_{7\mu}$         &  $10.7^2$        & $7.8\sim8.4$           & $10.14\pm0.10$    & 44.8  \\
                        & $J_{8\mu}$         &  $10.7^2$        & $7.8\sim8.4$           & $10.14\pm0.09$    & 44.8
\vspace{5pt} \\
                        & $J_{3\mu}$         &  $10.9^2$        & $8.5\sim9.5$           & $10.34\pm0.09$    & 46.1 \\
$sb\bar s\bar b$ system & $J_{4\mu}$         &  $10.8^2$        & $8.5\sim9.1$           & $10.25\pm0.10$    & 43.3 \\
                        & $J_{7\mu}$         &  $10.8^2$        & $7.5\sim8.6$           & $10.24\pm0.11$    & 47.1  \\
                        & $J_{8\mu}$         &  $10.8^2$        & $7.5\sim8.6$           & $10.24\pm0.10$    & 47.1  \\
\hlinewd{.8pt}
\end{tabular}
\tabcaption{The threshold value, Borel window, mass and pole
contribution corresponding to the currents with $J^{PC}=1^{++}$ in the
$qc\bar q\bar c$, $sc\bar s\bar c$, $qb\bar q\bar b$ and $sb\bar
s\bar b$ systems.\label{table4}}
\end{center}
\begin{center}
\begin{tabular}{cccccc}
\hlinewd{.8pt}
                   & Currents & $s_0(\mbox{GeV}^2)$&$[M^2_{\mbox{min}}$,$M^2_{\mbox{max}}](\mbox{GeV}^2)$&$m_X$\mbox{(GeV)}&PC(\%)\\
\hline
                        & $J_{3\mu}$         &  $4.6^2$            & $3.0\sim3.4$           & $4.16\pm0.10$     & 46.2  \\
$qc\bar q\bar c$ system & $J_{4\mu}$         &  $4.5^2$            & $3.0\sim3.3$           & $4.02\pm0.09$     & 44.6  \\
                        & $J_{5\mu}$         &  $4.5^2$            & $3.0\sim3.4$           & $4.00\pm0.11$     & 46.0  \\
                        & $J_{6\mu}$         &  $4.6^2$            & $3.0\sim3.4$           & $4.14\pm0.09$     & 47.0
\vspace{5pt} \\
                        & $J_{3\mu}$         &  $4.7^2$            & $3.0\sim3.6$           & $4.24\pm0.10$     & 49.6  \\
$sc\bar s\bar c$ system & $J_{4\mu}$         &  $4.6^2$            & $3.0\sim3.5$           & $4.12\pm0.11$     & 47.3  \\
                        & $J_{5\mu}$         &  $4.5^2$            & $3.0\sim3.3$           & $4.03\pm0.11$     & 44.2  \\
                        & $J_{6\mu}$         &  $4.6^2$            & $3.0\sim3.4$           & $4.16\pm0.11$     & 46.0
\vspace{5pt} \\
                        & $J_{3\mu}$         &  $10.6^2$           & $7.5\sim8.5$           & $10.08\pm0.10$    & 45.9  \\
$qb\bar q\bar b$ system & $J_{4\mu}$         &  $10.6^2$           & $7.5\sim8.5$           & $10.07\pm0.10$    & 46.2  \\
                        & $J_{5\mu}$         &  $10.6^2$           & $7.5\sim8.4$           & $10.05\pm0.10$    & 45.3  \\
                        & $J_{6\mu}$         &  $10.7^2$           & $7.5\sim8.7$           & $10.15\pm0.10$    & 47.6
\vspace{5pt} \\
                        & $J_{3\mu}$         &  $10.6^2$           & $7.5\sim8.3$           & $10.11\pm0.10$    & 43.8  \\
$sb\bar s\bar b$ system & $J_{4\mu}$         &  $10.6^2$           & $7.5\sim8.4$           & $10.10\pm0.10$    & 44.1  \\
                        & $J_{5\mu}$         &  $10.6^2$           & $7.5\sim8.3$           & $10.08\pm0.10$    & 43.7  \\
                        & $J_{6\mu}$         &  $10.7^2$           & $7.5\sim8.5$           & $10.18\pm0.10$    & 46.5  \\
\hlinewd{.8pt}
\end{tabular}
\tabcaption{The threshold value, Borel window, mass and pole
contribution corresponding to the currents with $J^{PC}=1^{+-}$ in the
$qc\bar q\bar c$, $sc\bar s\bar c$, $qb\bar q\bar b$ and $sb\bar
s\bar b$ systems.\label{table5}}
\end{center}
\section{Decay Patterns of the Charmonium-like States}\label{sec:summary}

The decay properties of the charmonium-like states are important
for the study of their structures and detections at experiments.
In this section, we study the decay patterns of the
charmonium-like states with $J^{PC}=1^{-+}, 1^{--}, 1^{++}$ and
$1^{+-}$. Only the two-body hadronic decay is considered.
Replacing the D meson by the B meson, one gets the decay patterns
of the bottomonium-like states so long as the kinematics allows.

The G-parity of a charmonium-like state is defined as
$G=C\cdot(-1)^I$, where $I$ is the isospin. By considering the
conservation of the angular momentum, P-parity, C-parity, isospin
and G-parity, we collect the S-wave and P-wave decay modes of
these charmonium-like states in Table~\ref{table6} and
Table~\ref{table7}. For the vector channel, one notes that the
S-wave decay modes are dominant and the final states always
contain a S-wave and P-wave meson pair. Such a decay pattern is
also speculated to be characteristic of the hybrid meson. In fact,
the tetraquark state mixes easily with the hybrid state $cG\bar
c$. In quantum field theory the charmonium-like tetraquark
operator and the $cG\bar c$ hybrid operator with the same quantum
numbers probably couple to the same physical state. In
Table~\ref{table7}, the S-wave decay modes $J/\psi\omega$ and
$J/\psi\rho$ are listed in the $1^{++}$ channel, which is
consistent with the decay properties of
$X(3872)$~\cite{2003-Choi-p262001-262001, 2005-Abe-p-a}. Up to
now, no experimental signals are observed for the charmonium-like
$1^{-+}$ and $1^{+-}$ states. $Y(4360)$ and $Y(4660)$ are only
observed in the $\psi(2S)\pi^+\pi^-$
channel~\cite{2007-Wang-p142002-142002,
2010-Nakamura-p75021-75021}. The possible decay modes listed in
Table~\ref{table6} and Table~\ref{table7} may be useful to the
future search of these interesting charmonium-like and
bottomonium-like states at the experimental facilities such as
Super-B factories, PANDE, LHC and RHIC.

\begin{center}
\begin{tabular}{ccc}
\hlinewd{.8pt}
        $I^GJ^{PC}$  &                     $S$-wave                         &                           $P$-wave                \\
\hline
               &    $D^{\ast}(2007)^0\bar{D}_0^{\ast}(2400)^0+c.c.$,  &  $D^0(1865)\bar{D}^0(1865)$, $D^{\ast}(2007)^0\bar{D}^{\ast}(2007)^0$, \\
  $0^-1^{--}$  &    $D_1(2420)^0\bar{D}^0(1865)+c.c.$,                &  $\chi_{c0}(1P) h_1(1170)$, $\chi_{c1}(1P) h_1(1170)$, \\
               &    $D_1(2420)^0\bar{D}^{\ast}(2007)^0+c.c.$,$J/\psi f_0(980)$, &$J/\psi\eta$, $\psi(2S)\eta^{\prime}$, $\psi(2S) \eta$, $\eta_c(1S)\omega, \eta_c(2S)\omega$\\
               &    $\psi(2S) f_0(980)$, $\chi_{c0}(1P)\omega$, $\chi_{c1}(1P)\omega$,   &
\vspace{5pt} \\
               &    $D^{\ast}(2007)^0\bar{D}_0^{\ast}(2400)^0+c.c.$,  &  $D^0(1865)\bar{D}^0(1865)$, $D^{\ast}(2007)^0\bar{D}^{\ast}(2007)^0$,\\
  $1^+1^{--}$  &    $D_1(2420)^0\bar{D}^0(1865)+c.c.$,                &  $\chi_{c0}(1P)b_1(1235)$, $\chi_{c1}(1P)b_1(1235)$, $J/\psi\pi$, \\
               &    $D_1(2420)^0\bar{D}^{\ast}(2007)^0+c.c.$, $\chi_{c0}(1P)\rho$,   &  $\psi(2S)\pi$, $\eta_c(1S)\rho, \eta_c(2S)\rho$  \\
               &    $\chi_{c1}(1P)\rho$, $a_1(1260)J/\psi$, $b_1(1235)\eta_c(1S)$,       &\\
\hline
               &    $D^{\ast}(2007)^0\bar{D}_0^{\ast}(2400)^0+c.c.$,   &  $D^0(1865)\bar{D}^0(1865)$, $D^{\ast}(2007)^0\bar{D}^{\ast}(2007)^0$, \\
  $0^+1^{-+}$  &    $D_1(2420)^0\bar{D}^0(1865)+c.c.$, &  $\chi_{c0}(1P)f_0(600), \chi_{c0}(1P)f_0(980)$,  \\
               &    $D_1(2420)^0\bar{D}^{\ast}(2007)^0+c.c.$, &  $\eta_c(1S)\eta, \eta_c(1S)\eta^{\prime}$, $J/\psi\omega$, $\psi(2S)\omega$\\
               &    $f_1(1285)\eta_c(1S)$, $\chi_{c1}(1P)\eta$, $\chi_{c1}(1P)\eta^{\prime}$  &  $\chi_{c1}(1P)f_0(600), \chi_{c1}(1P)f_0(980)$,
\vspace{5pt} \\
               &    $D^{\ast}(2007)^0\bar{D}_0^{\ast}(2400)^0+c.c.$, $a_1(1260)\eta_c(1S)$,  & $D^0(1865)\bar{D}^0(1865)$, $D^{\ast}(2007)^0\bar{D}^{\ast}(2007)^0$,  \\
  $1^-1^{-+}$  &    $D_1(2420)^0\bar{D}^0(1865)+c.c.$, $b_1(1235)J/\psi$,                    & $\chi_{c0}(1P)a_0(980)$, $\eta_c(1S)\pi, \eta_c(2S)\pi$,\\
               &    $D_1(2420)^0\bar{D}^{\ast}(2007)^0+c.c.$, $\chi_{c1}(1P)\pi$             & $J/\psi\rho$, $\psi(2S)\rho$\\
\hlinewd{.8pt}
\end{tabular}
\tabcaption{The possible decay modes of the $1^{--}$ and $1^{-+}$
charmonium-like states. \label{table6}}
\end{center}

\begin{center}
\begin{tabular}{ccc}
\hlinewd{.8pt}
   $I^GJ^{PC}$  &                     $S$-wave                         &                           $P$-wave                \\
\hline
         $0^+1^{++}$ &    $D^0(1865)\bar{D}^{\ast}(2007)^0+c.c.$, $\chi_{c1}(1P)f_0(600)$,  &  $\chi_{c0}(1P)\eta$, $\chi_{c1}(1P)\eta$, \\
               &    $D(1870)^+D^{\ast}(2010)^{-}+c.c.$, $J/\psi\omega$                &  $\eta_c(1S)f_0(600)$
\vspace{5pt} \\
         $1^-1^{++}$ &    $D^0(1865)\bar{D}^{\ast}(2007)^0+c.c.$,                           &  $\chi_{c0}(1P)\pi$,\\
               &    $D(1870)^+D^{\ast}(2010)^{-}+c.c.$, $J/\psi\rho$                  &  $\chi_{c1}(1P)\pi$, $\eta_c(1S)a_1(1260)$ \\
\hline
         $0^-1^{+-}$ &    $D^0(1865)\bar{D}^{\ast}(2007)^0+c.c.$,           &  $\eta_c(1S)h_1(1170)$, $J/\psi f_0(600)$ \\
               &    $D(1870)^+D^{\ast}(2010)^{-}+c.c.$,               & $\chi_{c0}(1P)\omega$ \\
               &    $J/\psi\eta$, $J/\psi\eta^{\prime}$, $\eta_c(1S)\omega$ &
\vspace{5pt} \\
         $1^+1^{+-}$ &    $D^0(1865)\bar{D}^{\ast}(2007)^0+c.c.$, $J/\psi\pi$, $\psi(2S)\pi$,  &  $\eta_c(1S)b_1(1235)$, $\chi_{c0}(1P)\rho$\\
               &    $D(1870)^+D^{\ast}(2010)^{-}+c.c.$, $\eta_c(1S)\rho$                 & \\
\hlinewd{.8pt}
\end{tabular}
\tabcaption{The possible decay modes of the $1^{++}$ and $1^{+-}$
charmonium-like states. \label{table7}}
\end{center}

\section{Summary}\label{sec:summary}

We have constructed a matrix $O$ which is composed of the
tetraquark operators with different Lorentz structures. The
charge-conjugation transformation of the matrix is equal to its
transpose. With this relation, we can define the symmetric matrix
$S$ and antisymmetric matrix $A$. Considering the color structure,
the elements of $S$ and $A$ are the tetraquark operators with
definite C-parities. Then we can obtain all the tetraquark
interpolating currents with $J^{PC}=1^{-+}, 1^{--}, 1^{++}$ and
$1^{+-}$, as shown in Eqs.~(\ref{currents1})-(\ref{currents2}).

At the hadronic level, there exists big difference between the
molecular states and tetraquark states. The molecular states are
commonly assumed to be bound states of two hadrons formed by the
exchange of the color-singlet light mesons while the tetraquark
states are generally bound by the QCD force at the quark gluon
level. However, within the framework of the QCD sum rule approach,
the only difference lies in the interpolating current used in the
study of the molecular and tetraquark states. Other procedures
including the operator product expansion, the calculation of the
Wilsion coefficient and numerical analysis are the same. In
principle, if we exhaust all the possible molecular-type currents
and all the possible tetraquark-type currents, we can rigorously
show that these two sets of interpolating currents are equivalent
by using the Fierz rearrangement.

However, there exists important difference between one single
molecular-type current and one single tetraquark-type current. For
example, every single tetraquark-type current is a linear
combination of several independent molecular-type currents. In
this respect, one well-known example is the light scalar-isoscalar
sigma meson. The tetraquark-type current (or their
combination/mixing) leads to a better result than the simple
pion-pion molecular current. It's possible to distinguish the
tetraquark and molecular structures after exhaustive and
comprehensive hard work, which is one of the motivations of our
present systematical investigation.

By studying the two-point correlation functions, we have
calculated the spectral densities of these currents at the
quark-gluon level. The four quark condensate $\qq^2$ is dominant
for all the currents with $J^{PC}=1^{-+}$ and $1^{--}$ in the
$qc\bar q\bar c$ systems. For the currents with $J^{PC}=1^{++}$
and $1^{+-}$, however, the quark condensate $\qq$ becomes the most
important corrections to the correlation functions. These
properties of the spectral densities lead to a better OPE
convergence for the currents in the vector channel than that in
the axial-vector channel. The $m_s$ related terms are kept in the
spectral densities in order to study the contribution of the
strange quark in the $sc\bar s\bar c$ system. Actually, they give
important corrections and enhance the stabilities of QCD sum
rules.

The tetraquark assignments of $X(3872)$ and $Y(4660)$
have been studied within the framework of the QCD sum rule
approach, as mentioned in the introduction. In
Ref.~\cite{2007-Matheus-p14005-14005}, the current $J_{4\mu}$ was
used to study the $1^{++}$ $qc\bar q\bar c$ system with the
extracted mass around 3.92 GeV. The current $J_{3\mu}$ was used to
study the $1^{--}$ $sc\bar s\bar c$ system with the extracted mass
around 4.65 GeV Ref.~\cite{2009-Albuquerque-p53-66}. One
difference of our present analysis and the previous ones lies in
the criteria of fixing the Borel window and the value of the
threshold parameter $s_0$, which leads to the slightly different
extracted masses of the states. We have imposed a strict
requirement that (1) the pole contribution be larger than 40\% and
(2)dual stability, i.e., the variation of the extracted mass with
both $s_0$ and the Borel parameter be minimum. The other
difference is that we have exhausted the tetraquark interpolating
currents.

In the working range of the Borel parameter, only the currents
$J_{1\mu}, J_{4\mu}$ and $J_{7\mu}$ with $J^{PC}=1^{--}$ display
stable QCD sum rules in the $qc\bar q\bar c$ system. The extracted
mass is around $4.6\sim 4.7$ GeV from these currents, which is
consistent with the mass of the meson $Y(4660)$. This result
implies a possible tetraquark interpretation for $Y(4660)$. In the
$sc\bar s\bar c$ system, all currents except $J_{5\mu}, J_{6\mu}$
have stable QCD sum rules. The mass difference
$m_{X^s}-m_X\approx0.2$ GeV for the same type of the interpolating
current, which is roughly $2(m_s-m_q)$. For the $1^{--}$
bottomonium-like states, the masses lie around $10.5$ GeV and
$10.6$ GeV for the $qb\bar q\bar b$ and $sb\bar s\bar b$ systems,
respectively. The Borel window for the currents in the
axial-vector channel is very small because of the bad OPE
convergence. For the currents with $J^{PC}=1^{++}$ in the $qc\bar
q\bar c$ system, only $J_{3\mu}$ and $J_{4\mu}$ lead to stable QCD
sum rules. The same situation occurs in the $sc\bar s\bar c$
system. The extracted masses are about $4.0\sim 4.2$ GeV, which is
$0.1\sim 0.2$ GeV higher than the mass of $X(3872)$.
The masses of the $1^{++}$ bottomonium-like states are about
$10.2$ GeV for both the $qb\bar q\bar b$ and $sb\bar s\bar b$ systems.

\section*{Acknowledgments}
The authors thank Professor W. Z. Deng for useful discussions.
This project was supported by the National Natural Science
Foundation of China under Grants 10625521,10721063 and Ministry of
Science and Technology of China(2009CB825200).



\appendix
\section{THE SPECTRAL DENSITIES}

In this appendix we show the spectral densities of the tetraquark
interpolating currents defined in Eq.~(\ref{currents1}). Various
power corrections include the four quark condensate $\qq^2$, quark
gluon mixed condensate $\qGq$ and dimension eight condensate
$\qGq\qq$:
\begin{eqnarray}
\rho(s)=\rho^{pert}(s)+\rho^{\qq}(s)+\rho^{\GGa}(s)+\rho^{\qq^2}(s)
+\rho^{\qGq}(s).
\end{eqnarray}
The integration limits in the expressions are:
\begin{eqnarray}
\nonumber
\alpha_{max}&=&\frac{1+\sqrt{1-4m_c^2/s}}{2},\hspace{1cm}
\alpha_{min}=\frac{1-\sqrt{1-4m_c^2/s}}{2}\non
\beta_{max}&=&1-\alpha,\hspace{2.8cm} \beta_{min}=\frac{\alpha
m_c^2}{\alpha s-m_c^2}.
\end{eqnarray}

\subsection{The spectral densities for the currents with $J^{PC}=1^{-+}$}
For the interpolating current $J_{1\mu}$:
\begin{eqnarray}
\nonumber
\rho^{pert}_1(s)&=&\frac{1}{3\times2^8\pi^6}\dab\frac{(1-\alpha-\beta)\f(s)^3}{\alpha^2\beta^3}
\bigg\{\frac{3(1+\alpha+\beta)\f(s)}{\alpha} \non
&&\hspace{4.5cm}+\frac{2m_c^2(1-\alpha-\beta)^2}{\alpha}+24m_cm_q(1-\alpha-\beta)\bigg\}
\, , \non \rho^{\qq}_1(s)&=&\frac{\qq}{8\pi^4}\dab \non
&&\frac{\f(s)}{\alpha\beta}\bigg\{\frac{2m_c(1-\alpha-\beta)\f(s)}{\beta}+m_q[(\alpha+\beta-3)m_c^2-4\alpha\beta
s]\bigg\} \, , \non
\rho^{\GGa}_1(s)&=&-\frac{\GGb}{3^2\times2^{12}\pi^6}\dab \non
&&\bigg\{\frac{(1-\alpha-\beta)^2m_c^2}{\alpha^2}\bigg[\frac{[96\beta^2+36\alpha\beta-(1-\alpha-\beta)(5\alpha+48\beta)]
\f(s)}{\alpha\beta^2}+\frac{30\f(s)}{\beta^2} \non
&&-\frac{96[2(\alpha+\beta)m_c^2-3\alpha\beta
s]+16(1-\alpha-\beta)m_c^2}{\alpha}\bigg]-\frac{48(1-\alpha-\beta)\f(s)s}{\alpha}\bigg\}
\, , \non
\rho^{\qGq}_1(s)&=&\frac{m_q\qGq(8m_c^2+s)}{96\pi^4}\sqrt{1-4m_c^2/s}
\non
&&-\frac{\qGq}{3\times2^7\pi^4}\dab\bigg\{\frac{m_c[48\alpha\beta+(1-\alpha-\beta)(7\alpha+6\beta)]\f(s)}{\alpha\beta^2}
\non
&&\hspace{4.5cm}-\frac{2m_c(1-\alpha-\beta)[2(\alpha+\beta)m_c^2-3\alpha\beta
s]}{\beta^2}+\frac{6m_q[(\alpha+\beta-2)m_c^2-\alpha\beta
s]}{\alpha}\bigg\} \, , \non
\rho^{\qq^2}_1(s)&=&-\frac{\qq^2(8m_c^2-6m_cm_q+s)}{36\pi^2}\sqrt{1-4m_c^2/s}
\, ,
\end{eqnarray}
\begin{eqnarray}
\Pi^{\qGq\qq}_1(M_B^2)&=&-\frac{\qGq\qq}{12\pi^2}\int_0^1d\alpha\bigg\{(M_B^2+\frac{m_c^2}{\alpha})(1-\alpha)
+\frac{2m_c^4}{\alpha^2M_B^2}+\frac{m_c^2}{2\alpha}+\frac{M_B^2\alpha}{4}\bigg\}\efun
\, .
\end{eqnarray}

For the interpolating current $J_{2\mu}$:
\begin{eqnarray}
\nonumber \rho^{pert}_2(s)&=&\frac{3}{2}\rho^{pert}_1(s),~~~~~
\rho^{\qq}_2(s)=\frac{3}{2}\rho^{\qq}_1(s),~~~~~
\rho^{\qq^2}_2(s)=\frac{3}{2}\rho^{\qq^2}_1(s) \, , \non
\rho^{\GGa}_2(s)&=&\frac{\GGb}{3^2\times2^{12}\pi^6}\dab \non
&&\bigg\{
\frac{(1-\alpha-\beta)^2m_c^2}{\alpha^2}\bigg[\frac{(1-\alpha-\beta)(13\alpha+72\beta)\f(s)}{\alpha\beta^2}-\frac{42\f(s)}{\beta^2}
\non &&+\frac{24[(1+5\alpha+5\beta)m_c^2-12\alpha\beta
s]}{\alpha}\bigg]+\frac{12\f(s)}{\alpha\beta}[(5-\alpha-\beta)m_c^2-(1-\alpha-\beta)^2s-2\alpha\beta
s]\bigg\} \, , \non
\rho^{\qGq}_2(s)&=&\frac{m_q\qGq(8m_c^2+s)}{64\pi^4}\sqrt{1-4m_c^2/s}
\non
&&-\frac{m_c\qGq}{2^7\pi^4}\dab\bigg\{\frac{3(1-\alpha-\beta)[3(\alpha+\beta)m_c^2-5\alpha\beta
s]}{\beta^2}+\frac{20\f(s)}{\beta}\bigg\} \, ,
\end{eqnarray}
\begin{eqnarray}
\Pi^{\qGq\qq}_2(M_B^2)&=&-\frac{\qGq\qq}{8\pi^2}\int_0^1d\alpha\bigg\{M_B^2(1-\alpha)
+\frac{1-\alpha}{\alpha}m_c^2+\frac{2m_c^4}{\alpha^2M_B^2}\bigg\}\efun
\, .
\end{eqnarray}

For the interpolating current $J_{3\mu}$:
\begin{eqnarray}
\nonumber \rho^{pert}_3(s)&=&\frac{1}{2}\rho^{pert}_1(s),~~~~~
\rho^{\qq}_3(s)=\frac{1}{2}\rho^{\qq}_1(s),~~~~~
\rho^{\qq^2}_3(s)=\frac{1}{2}\rho^{\qq^2}_1(s) \, , \non
\rho^{\GGa}_3(s)&=&\frac{\GGb}{3^2\times2^{12}\pi^6}\dab \non
&&\bigg\{
\frac{(1-\alpha-\beta)^2m_c^2}{\alpha^2}\bigg[\frac{[36\alpha\beta+24\beta(1-\alpha-\beta)
-\alpha(5+\alpha+\beta)]\f(s)}{\alpha\beta^2}+\frac{48[(\alpha+\beta)m_c^2-2\alpha\beta
s]}{\alpha} \non
&&+\frac{8(1-\alpha-\beta)m_c^2}{\alpha}\bigg]-\frac{48(1-\alpha-\beta)\f(s)s}{\alpha}\bigg\}
\, , \non
\rho^{\qGq}_3(s)&=&\frac{m_q\qGq(8m_c^2+s)}{192\pi^4}\sqrt{1-4m_c^2/s}
\non
&&+\frac{\qGq}{3\times2^7\pi^4}\dab\bigg\{\frac{6m_c(1-5\alpha-\beta)\f(s)}{\alpha\beta}
\non
&&\hspace{4.5cm}-\frac{m_c(1-\alpha-\beta)[15(\alpha+\beta)m_c^2-29\alpha\beta
s]}{\beta^2}+\frac{6m_q[(\alpha+\beta-2)m_c^2-\alpha\beta
s]}{\alpha}\bigg\} \, ,
\end{eqnarray}
\begin{eqnarray}
\Pi^{\qGq\qq}_3(M_B^2)&=&-\frac{\qGq\qq}{48\pi^2}\int_0^1d\alpha\bigg\{(2-3\alpha)M_B^2-2m_c^2
+\frac{4m_c^4}{\alpha^2M_B^2}\bigg\}\efun \, .
\end{eqnarray}

For the interpolating current $J_{4\mu}$:
\begin{eqnarray}
\nonumber \rho^{pert}_4(s)&=&3\rho^{pert}_1(s),~~~~~
\rho^{\qq}_4(s)=3\rho^{\qq}_1(s),~~~~~
\rho^{\qq^2}_4(s)=3\rho^{\qq^2}_1(s) \, , \non
\rho^{\GGa}_4(s)&=&\frac{\GGb}{3^2\times2^{12}\pi^6}\dab\bigg\{\frac{(1-\alpha-\beta)^2m_c^2}{\alpha^2}
\bigg[\frac{48[m_c^2(1+5\alpha+5\beta)-12\alpha\beta s]}{\alpha}
\non
&&+\frac{[216\alpha\beta+(1-\alpha-\beta)(65\alpha+144\beta)]\f(s)}{\alpha\beta^2}-\frac{210\f(s)}{\beta^2}\bigg]
\non
&&+\frac{12\f(s)}{\alpha\beta}[5(5-\alpha-\beta)m_c^2-5(1-\alpha-\beta)^2s-24(1-\alpha-\beta)\beta
s-10\alpha\beta s]\bigg\} \, , \non
\rho^{\qGq}_4(s)&=&\frac{m_q\qGq(8m_c^2+s)}{32\pi^4}\sqrt{1-4m_c^2/s}
\non
&&-\frac{\qGq}{2^7\pi^4}\dab\bigg\{\frac{m_c[28\alpha\beta-(23\alpha+12\beta)(1-\alpha-\beta)]\f(s)}{\alpha\beta^2}
\non
&&+\frac{22m_c(1-\alpha-\beta)[2(\alpha+\beta)m_c^2-3\alpha\beta
s]}{\beta^2}-\frac{12m_q[(\alpha+\beta-2)m_c^2-\alpha\beta
s]}{\alpha}\bigg\} \, ,
\end{eqnarray}
\begin{eqnarray}
\Pi^{\qGq\qq}_4(M_B^2)&=&-\frac{\qGq\qq}{8\pi^2}\int_0^1d\alpha\bigg\{M_B^2(2-3\alpha)-2m_c^2+\frac{4m_c^4}{\alpha^2M_B^2}
\bigg\}\efun \, .
\end{eqnarray}

From these results the expressions for the currents $J_{5\mu},
J_{6\mu}, J_{7\mu}$ and $J_{8\mu}$ can then be obtained conveniently
by the replacement $m_c\rightarrow -m_c$:
\begin{eqnarray}
\rho_1(s)\xrightarrow{m_c\rightarrow -m_c}\rho_5(s),
\rho_2(s)\xrightarrow{m_c\rightarrow -m_c}\rho_8(s),
\rho_3(s)\xrightarrow{m_c\rightarrow -m_c}\rho_7(s),
\rho_4(s)\xrightarrow{m_c\rightarrow -m_c}\rho_6(s) \, .
\end{eqnarray}

\subsection{The spectral densities for the currents with $J^{PC}=1^{--}$}
For the interpolating current $J_{1\mu}$:
\begin{eqnarray}
\nonumber \rho^{pert}_1(s)&=&\frac{1}{3\times2^8\pi^6}\dab \non
&&\frac{(1-\alpha-\beta)\f(s)^3}{\alpha^3\beta^3}\bigg\{3(1+\alpha+\beta)\f(s)-2m_c^2(1-\alpha-\beta)^2\bigg\}
\, , \non
\rho^{\qq}_1(s)&=&-\frac{m_q\qq}{8\pi^4}\dab\frac{\f(s)[(5-\alpha-\beta)m_c^2+2\alpha\beta
s]}{\alpha\beta} \, , \non
\rho^{\GGa}_1(s)&=&-\frac{\GGb}{3^2\times2^{12}\pi^6}\dab \non
&&\bigg\{
\frac{(1-\alpha-\beta)^2\f(s)m_c^2}{\alpha^2\beta}\bigg[\frac{48(1-\alpha-\beta)}{\alpha}-\frac{5(5+\alpha+\beta)}{\beta}-36\bigg]
\non
&&+\frac{16(1-\alpha-\beta)^2[(1-7\alpha-7\beta)m_c^2+12\alpha\beta
s]m_c^2}{\alpha^3}-\frac{48(1-\alpha-\beta)\f(s)s}{\alpha}\bigg\}
\, , \non
\rho^{\qGq}_1(s)&=&\frac{m_q\qGq(16m_c^2-s)}{3\times2^5\pi^4}\sqrt{1-4m_c^2/s}+\frac{\qGq}{3\times2^7\pi^4}\dab
\non
&&\bigg\{\frac{(1-\alpha-\beta)m_c}{\beta}\bigg[\frac{6\f(s)}{\alpha}+\frac{27(\alpha+\beta)m_c^2-49\alpha\beta
s}{\beta}\bigg]+\frac{6m_q((2+\alpha+\beta)m_c^2-\alpha\beta
s)}{\alpha}\bigg\} \, , \non
\rho^{\qq^2}_1(s)&=&-\frac{\qq^2(16m_c^2-s)}{36\pi^2}\sqrt{1-4m_c^2/s}
\, ,
\end{eqnarray}
\begin{eqnarray}
\Pi^{\qGq\qq}_1(M_B^2)&=&\frac{\qGq\qq}{48\pi^2}\int_0^1d\alpha\bigg\{(4-3\alpha)M_B^2
+\frac{2(1-2\alpha)m_c^2}{\alpha}-\frac{8m_c^4}{\alpha^2M_B^2}\bigg\}\efun
\, .
\end{eqnarray}

For the interpolating current $J_{2\mu}$:
\begin{eqnarray}
\nonumber \rho^{pert}_2(s)&=&\frac{3}{2}\rho^{pert}_1(s),~~~~~
\rho^{\qq}_2(s)=\frac{3}{2}\rho^{\qq}_1(s),~~~~~
\rho^{\qq^2}_2(s)=\frac{3}{2}\rho^{\qq^2}_1(s) \, , \non
\rho^{\GGa}_2(s)&=&-\frac{\GGb}{3^2\times2^{12}\pi^6}\dab \non
&&\bigg\{
\frac{(1-\alpha-\beta)^2\f(s)}{\alpha^2\beta}\bigg[\frac{72(1-\alpha-\beta)m_c^2}{\alpha}
-\frac{(29+13\alpha+13\beta)m_c^2-12\alpha\beta s}{\beta}\bigg]
\non &&+\frac{12\f(s)[(5-\alpha-\beta)m_c^2+2\alpha\beta
s]}{\alpha\beta}
+\frac{24(1-\alpha-\beta)^2[(1-7\alpha-7\beta)m_c^2+12\alpha\beta
s]m_c^2}{\alpha^3}\bigg\} \, , \non
\rho^{\qGq}_2(s)&=&\frac{m_q\qGq(16m_c^2-s)}{64\pi^4}\sqrt{1-4m_c^2/s}+\frac{m_c\qGq}{3\times2^7\pi^4}\dab
\non
&&\frac{(1-\alpha-\beta)}{\beta}\bigg[\frac{12\f(s)}{\alpha}+\frac{3(\alpha+\beta)m_c^2-13\alpha\beta
s}{\beta}\bigg] \, ,
\end{eqnarray}
\begin{eqnarray}
\Pi^{\qGq\qq}_2(M_B^2)&=&\frac{\qGq\qq}{8\pi^2}\int_0^1d\alpha\bigg\{(1-\alpha)M_B^2
+\frac{(1-\alpha)m_c^2}{\alpha}-\frac{2m_c^4}{\alpha^2M_B^2}\bigg\}\efun
\, .
\end{eqnarray}

For the interpolating current $J_{3\mu}$:
\begin{eqnarray}
\nonumber \rho^{pert}_3(s)&=&\frac{1}{2}\rho^{pert}_1(s),~~~~~
\rho^{\qq}_3(s)=\frac{1}{2}\rho^{\qq}_1(s),~~~~~
\rho^{\qq^2}_3(s)=\frac{1}{2}\rho^{\qq^2}_1(s) \, , \non
\rho^{\GGa}_3(s)&=&-\frac{\GGb}{3^2\times2^{12}\pi^6}\dab \non
&&\bigg\{
\frac{(1-\alpha-\beta)^2\f(s)m_c^2}{\alpha^2\beta}\bigg[\frac{24(1-\alpha-\beta)}{\alpha}-\frac{5+\alpha+\beta}{\beta}+36\bigg]
\non
&&+\frac{8(1-\alpha-\beta)^2[(1-7\alpha-7\beta)m_c^2+12\alpha\beta
s]m_c^2}{\alpha^3}+\frac{48(1-\alpha-\beta)\f(s)s}{\alpha}\bigg\}
\, , \non
\rho^{\qGq}_3(s)&=&\frac{m_q\qGq(16m_c^2-s)}{3\times2^6\pi^4}\sqrt{1-4m_c^2/s}-\frac{\qGq}{3\times2^7\pi^4}\dab
\non
&&\bigg\{\frac{(1-\alpha-\beta)m_c}{\beta}\bigg[\frac{6\f(s)}{\alpha}+\frac{9(\alpha+\beta)m_c^2-19\alpha\beta
s}{\beta}\bigg]+\frac{6m_q((2+\alpha+\beta)m_c^2-\alpha\beta
s)}{\alpha}\bigg\} \, , \non
\end{eqnarray}
\begin{eqnarray}
\Pi^{\qGq\qq}_3(M_B^2)&=&\frac{\qGq\qq}{48\pi^2}\int_0^1d\alpha\bigg\{(2-3\alpha)M_B^2
+\frac{2(2-\alpha)m_c^2}{\alpha}-\frac{4m_c^4}{\alpha^2M_B^2}\bigg\}\efun
\, .
\end{eqnarray}

For the interpolating current $J_{4\mu}$:
\begin{eqnarray}
\nonumber \rho^{pert}_4(s)&=&3\rho^{pert}_1(s),~~~~~
\rho^{\qq}_4(s)=3\rho^{\qq}_1(s),~~~~~
\rho^{\qq^2}_4(s)=3\rho^{\qq^2}_1(s) \, , \non
\rho^{\GGa}_4(s)&=&-\frac{\GGb}{3^2\times2^{12}\pi^6}\dab \non
&&\bigg\{
\frac{(1-\alpha-\beta)^2\f(s)m_c^2}{\alpha^2\beta}\bigg[\frac{144(1-\alpha-\beta)}{\alpha}
+\frac{65(1-\alpha-\beta)}{\beta}-\frac{210}{\beta}+216\bigg] \non
&&+\frac{60(5-\alpha-\beta)\f(s)m_c^2}{\alpha\beta}+\frac{48(1-\alpha-\beta)^2[(1-7\alpha-7\beta)m_c^2+12\alpha\beta
s]m_c^2}{\alpha^3} \non
&&+\bigg[\frac{288(1-\alpha-\beta)}{\alpha}+\frac{60(1-\alpha-\beta)^2}{\alpha\beta}+120\bigg]\f(s)s\bigg\}
\, , \non
\rho^{\qGq}_4(s)&=&\frac{m_q\qGq(16m_c^2-s)}{32\pi^4}\sqrt{1-4m_c^2/s}+\frac{\qGq}{3\times2^7\pi^4}\dab
\non
&&\bigg\{\frac{(1-\alpha-\beta)m_c}{\beta}\bigg[\frac{24\f(s)}{\alpha}+\frac{87(\alpha+\beta)m_c^2-161\alpha\beta
s}{\beta}\bigg]-\frac{36m_q((2+\alpha+\beta)m_c^2-\alpha\beta
s)}{\alpha}\bigg\} \, , \non
\end{eqnarray}
\begin{eqnarray}
\Pi^{\qGq\qq}_4(M_B^2)&=&\frac{\qGq\qq}{8\pi^2}\int_0^1d\alpha\bigg\{(2-3\alpha)M_B^2
+\frac{2(2-\alpha)m_c^2}{\alpha}-\frac{4m_c^4}{\alpha^2M_B^2}\bigg\}\efun
\, .
\end{eqnarray}

From these results the expressions for the currents $J_{5\mu},
J_{6\mu}, J_{7\mu}$ and $J_{8\mu}$ can then be obtained conveniently
by the replacement $m_c\rightarrow -m_c$:
\begin{eqnarray}
\rho_1(s)\xrightarrow{m_c\rightarrow -m_c}\rho_5(s),
\rho_2(s)\xrightarrow{m_c\rightarrow -m_c}\rho_8(s),
\rho_3(s)\xrightarrow{m_c\rightarrow -m_c}\rho_7(s),
\rho_4(s)\xrightarrow{m_c\rightarrow -m_c}\rho_6(s) \, .
\end{eqnarray}

\subsection{The spectral densities for the currents with $J^{PC}=1^{++}$}
For the interpolating current $J_{1\mu}$:
\begin{eqnarray}
\nonumber
\rho^{pert}_1(s)&=&\frac{1}{2^8\pi^6}\dab\frac{(1-\alpha-\beta)\f(s)^2}{\alpha^2\beta^3}
\non
&&\bigg\{\frac{(1+\alpha+\beta)\f(s)^2}{\alpha}+12m_cm_q(1-\alpha-\beta)[(\alpha+\beta)m_c^2-3\alpha\beta
s]\bigg\} \, , \non \rho^{\qq}_1(s)&=&\frac{\qq}{8\pi^4}\dab \non
&&\frac{\f(s)}{\alpha\beta}\bigg\{\frac{m_c(1-\alpha-\beta)[3(\alpha+\beta)m_c^2-7\alpha\beta
s]}{\beta}+m_q[(4+\alpha+\beta)m_c^2-3\alpha\beta s]\bigg\} \, ,
\non \rho^{\GGa}_1(s)&=&\frac{\GGb}{3^2\times2^{12}\pi^6}\dab \non
&&\bigg\{\frac{(1-\alpha-\beta)^2m_c^2}{\alpha^2}\bigg[\frac{96[(\alpha+\beta)m_c^2-2\alpha\beta
s]}{\alpha} +\frac{5(5+\alpha+7\beta)\f(s)}{\beta^2}\bigg] \non
&&-\frac{6\f(s)}{\alpha\beta}\bigg[\frac{4(1-\alpha-\beta)(5m_c^2-2\alpha\beta
s)}{\alpha} -5m_c^2(1+\alpha+\beta)\bigg]\bigg\} \, , \non
\rho^{\qGq}_1(s)&=&-\frac{m_c^2m_q\qGq}{8\pi^4}\sqrt{1-4m_c^2/s}+\frac{\qGq}{3\times2^7\pi^4}\dab
\non
&&\bigg\{\frac{m_c(1-\alpha-\beta)[3(\alpha+\beta)m_c^2-\alpha\beta
s]}{\beta^2}-\frac{12m_q
m_c^2}{\beta}-\frac{(29m_c+5m_q)[3(\alpha+\beta)m_c^2-5\alpha\beta
s]}{\beta}\bigg\} \, , \non
\rho^{\qq^2}_1(s)&=&\frac{m_c\qq^2}{12\pi^2}(4m_c+2m_q-\frac{m_q
s}{4m_c^2-s})\sqrt{1-4m_c^2/s} \, ,
\end{eqnarray}
\begin{eqnarray}
\Pi^{\qGq\qq}_1(M_B^2)&=&-\frac{\qGq\qq}{3^2\times2^5\pi^2}\int_0^1d\alpha\bigg\{15\alpha
M_B^2-\frac{2(6-11\alpha)m_c^2}{\alpha(1-\alpha)}-\frac{48m_c^4}{\alpha^2M_B^2}\bigg\}\efun
\, .
\end{eqnarray}

For the interpolating current $J_{2\mu}$:
\begin{eqnarray}
\nonumber \rho^{pert}_2(s)&=&\frac{1}{2}\rho^{pert}_1(s),~~~~~
\rho^{\qq}_2(s)=\frac{1}{2}\rho^{\qq}_1(s),~~~~~
\rho^{\qq^2}_2(s)=\frac{1}{2}\rho^{\qq^2}_1(s) \, , \non
\rho^{\GGa}_2(s)&=&\frac{\GGb}{3^2\times2^{12}\pi^6}\dab \non
&&\bigg\{\frac{(1-\alpha-\beta)^2m_c^2}{\alpha^2}\bigg[\frac{48[(\alpha+\beta)m_c^2-2\alpha\beta
s]}{\alpha} +\frac{(5+\alpha+7\beta)\f(s)}{\beta^2}\bigg] \non
&&-\frac{6\f(s)}{\alpha\beta}\bigg[\frac{4(1-\alpha-\beta)(m_c^2+2\alpha\beta
s)}{\alpha} -m_c^2(1+\alpha+\beta)\bigg]\bigg\} \, , \non
\rho^{\qGq}_2(s)&=&-\frac{m_c^2m_q\qGq}{16\pi^4}\sqrt{1-4m_c^2/s}+\frac{\qGq}{3\times2^7\pi^4}\dab
\non
&&\bigg\{\frac{m_c(1-\alpha-\beta)[15(\alpha+\beta)m_c^2-29\alpha\beta
s]}{\beta^2}+\frac{12m_q
m_c^2}{\beta}-\frac{(13m_c+m_q)[3(\alpha+\beta)m_c^2-5\alpha\beta
s]}{\beta}\bigg\} \, ,
\end{eqnarray}
\begin{eqnarray}
\Pi^{\qGq\qq}_2(M_B^2)&=&-\frac{\qGq\qq}{3^2\times2^5\pi^2}\int_0^1d\alpha\bigg\{3\alpha
M_B^2+\frac{2(6-5\alpha)m_c^2}{\alpha(1-\alpha)}-\frac{24m_c^4}{\alpha^2M_B^2}\bigg\}\efun
\, .
\end{eqnarray}

For the interpolating current $J_{5\mu}$:
\begin{eqnarray}
\nonumber
\rho^{pert}_5(s)&=&\frac{1}{3\times2^8\pi^6}\dab\frac{(1-\alpha-\beta)\f(s)^2}{\alpha^2\beta^3}
\bigg\{\frac{9(1+\alpha+\beta)\f(s)^2}{\alpha} \non
&&+\frac{4m_c^2(1-\alpha-\beta)(5+\alpha+\beta)\f(s)}{\alpha}
+36m_cm_q(1-\alpha-\beta)[(\alpha+\beta)m_c^2-3\alpha\beta
s]\bigg\} \, , \non \rho^{\qq}_5(s)&=&\frac{\qq}{8\pi^4}\dab \non
&&\frac{\f(s)}{\alpha\beta}\bigg\{\frac{m_c(1-\alpha-\beta)[3(\alpha+\beta)m_c^2-7\alpha\beta
s]}{\beta}+m_q[(14+11\alpha+11\beta)m_c^2-23\alpha\beta s]\bigg\}
\, , \non
\rho^{\GGa}_5(s)&=&\frac{\GGb}{3^2\times2^{12}\pi^6}\dab\bigg\{\frac{\f(s)m_c^2}{\alpha\beta}
\bigg[\frac{96(1-\alpha-\beta)^2(5+\alpha+\beta)}{\alpha^2} \non
&&+\frac{54(1-\alpha-\beta)(3+\alpha+\beta)}{\alpha}+\frac{5(1-\alpha-\beta)^2(29+13\alpha+13\beta)}{\alpha\beta}
+90(1+\alpha+\beta)\bigg] \non
&&-\frac{\f(s)s}{\alpha\beta}\bigg[288(1-\alpha-\beta)\alpha+60(1-\alpha-\beta)^2+120\alpha\beta\bigg]
\non
&&+\frac{32(1-\alpha-\beta)^2[5(1+2\alpha+2\beta)m_c^2-18\alpha\beta
s]m_c^2}{\alpha^3}\bigg\} \, , \non
\rho^{\qGq}_5(s)&=&\frac{m_q\qGq(s-10m_c^2)}{48\pi^4}\sqrt{1-4m_c^2/s}-\frac{\qGq}{3\times2^7\pi^4}\dab
\non &&\bigg\{\frac{(29m_c-9m_q)[3(\alpha+\beta)m_c^2-5\alpha\beta
s]}{\beta}
+\frac{6m_c(1-\alpha-\beta)[3(\alpha+\beta)m_c^2-5\alpha\beta
s]}{\alpha\beta} \non
&&-\frac{(1-\alpha-\beta)[63(\alpha+\beta)m_c^2-129\alpha\beta
s]}{\beta^2} -\frac{72m_qm_c^2}{\beta}\bigg\} \, ,
\\
\rho^{\qq^2}_5(s)&=&\frac{\qq^2}{36\pi^2}\bigg[20m_c^2-2s+\frac{3m_qm_c(8m_c^2-3s)}{4m_c^2-s}\bigg]\sqrt{1-4m_c^2/s}
\, ,
\end{eqnarray}
\begin{eqnarray}
\nonumber
\Pi^{\qGq\qq}_5(M_B^2)&=&-\frac{\qGq\qq}{96\pi^2}\int_0^1d\alpha\bigg\{(7\alpha+16)
M_B^2-\frac{2(8\alpha^2-\alpha-20)m_c^2}{\alpha(1-\alpha)}-\frac{16(3-5\alpha)m_c^4}{\alpha^2(1-\alpha)M_B^2}\bigg\}\efun
\, .
\end{eqnarray}

For the interpolating current $J_{6\mu}$:
\begin{eqnarray}
\nonumber \rho^{pert}_6(s)&=&\frac{1}{2}\rho^{pert}_5(s),~~~~~
\rho^{\qq}_6(s)=\frac{1}{2}\rho^{\qq}_5(s),~~~~~
\rho^{\qq^2}_6(s)=\frac{1}{2}\rho^{\qq^2}_5(s) \, , \non
\rho^{\GGa}_6(s)&=&\frac{\GGb}{3^2\times2^{12}\pi^6}\dab\bigg\{\frac{\f(s)m_c^2}{\alpha\beta}
\bigg[\frac{48(1-\alpha-\beta)^2(5+\alpha+\beta)}{\alpha^2} \non
&&-\frac{18(1-\alpha-\beta)(3+\alpha+\beta)}{\alpha}+\frac{(1-\alpha-\beta)^2(29+13\alpha+13\beta)}{\alpha\beta}
+18(1+\alpha+\beta)\bigg] \non
&&-\frac{12[(1-\alpha-\beta)^2+2\alpha\beta]\f(s)s}{\alpha\beta}
+\frac{16(1-\alpha-\beta)^2[5(1+2\alpha+2\beta)m_c^2-18\alpha\beta
s]m_c^2}{\alpha^3}\bigg\} \, , \non
\rho^{\qGq}_6(s)&=&\frac{m_q\qGq(s-10m_c^2)}{96\pi^4}\sqrt{1-4m_c^2/s}-\frac{\qGq}{3\times2^7\pi^4}\dab
\\
&&\bigg\{\frac{(13m_c+3m_q)[3(\alpha+\beta)m_c^2-5\alpha\beta
s]}{\beta}
+\frac{(1-\alpha-\beta)(6\beta-9\alpha)[3(\alpha+\beta)m_c^2-5\alpha\beta
s]m_c}{\alpha\beta^2}\bigg\} \, , \non
\Pi^{\qGq\qq}_6&(M_B^2)&=-\frac{\qGq\qq}{96\pi^2}\int_0^1d\alpha\bigg\{(8\alpha+11)
M_B^2-\frac{2(4\alpha^2-9\alpha-4)m_c^2}{\alpha(1-\alpha)}-\frac{8(3-5\alpha)m_c^4}{\alpha^2(1-\alpha)M_B^2}\bigg\}\efun
\, .
\end{eqnarray}

From these results the expressions for the currents $J_{3\mu},
J_{4\mu}, J_{7\mu}$ and $J_{8\mu}$ can then be obtained conveniently
by the replacement $m_c\rightarrow -m_c$:
\begin{eqnarray}
\rho_1(s)\xrightarrow{m_c\rightarrow -m_c}\rho_3(s),
\rho_2(s)\xrightarrow{m_c\rightarrow -m_c}\rho_4(s),
\rho_5(s)\xrightarrow{m_c\rightarrow -m_c}\rho_7(s),
\rho_6(s)\xrightarrow{m_c\rightarrow -m_c}\rho_8(s) \, .
\end{eqnarray}

\subsection{The spectral densities for the currents with $J^{PC}=1^{+-}$}
For the interpolating current $J_{1\mu}$:
\begin{eqnarray}
\nonumber
\rho^{pert}_1(s)&=&\frac{1}{2^8\pi^6}\dab\frac{(1-\alpha-\beta)\f(s)^2}{\alpha^2\beta^3}
\non
&&\bigg\{\frac{(1+\alpha+\beta)\f(s)^2}{\alpha}+12m_cm_q(1-\alpha-\beta)[(\alpha+\beta)m_c^2-3\alpha\beta
s]\bigg\} \, , \non \rho^{\qq}_1(s)&=&\frac{\qq}{8\pi^4}\dab \non
&&\frac{\f(s)}{\alpha\beta}\bigg\{\frac{m_c(1-\alpha-\beta)[3(\alpha+\beta)m_c^2-7\alpha\beta
s]}{\beta}+m_q[(4+\alpha+\beta)m_c^2-3\alpha\beta s]\bigg\} \, ,
\non \rho^{\GGa}_1(s)&=&\frac{\GGb}{3^2\times2^{12}\pi^6}\dab \non
&&\bigg\{\frac{(1-\alpha-\beta)^2m_c^2}{\alpha^2}\bigg[\frac{96[(\alpha+\beta)m_c^2-2\alpha\beta
s]}{\alpha} -\frac{5(5+\alpha+7\beta)\f(s)}{\beta^2}\bigg] \non
&&+\frac{6\f(s)}{\alpha\beta}\bigg[\frac{4(1-\alpha-\beta)(5m_c^2+2\alpha\beta
s)}{\alpha} -5m_c^2(1+\alpha+\beta)\bigg]\bigg\} \, , \non
\rho^{\qGq}_1(s)&=&-\frac{m_c^2m_q\qGq}{8\pi^4}\sqrt{1-4m_c^2/s}-\frac{\qGq}{3\times2^7\pi^4}\dab
\non
&&\bigg\{\frac{m_c(1-\alpha-\beta)[27(\alpha+\beta)m_c^2-49\alpha\beta
s]}{\beta^2}+\frac{12m_q
m_c^2}{\beta}+\frac{(19m_c-5m_q)[3(\alpha+\beta)m_c^2-5\alpha\beta
s]}{\beta}\bigg\} \, , \non
\rho^{\qq^2}_1(s)&=&\frac{m_c\qq^2}{12\pi^2}(4m_c+2m_q-\frac{m_q
s}{4m_c^2-s})\sqrt{1-4m_c^2/s} \, ,
\end{eqnarray}
\begin{eqnarray}
\Pi^{\qGq\qq}_1(M_B^2)&=&\frac{\qGq\qq}{3^2\times2^5\pi^2}\int_0^1d\alpha\bigg\{15\alpha
M_B^2+\frac{2(6-\alpha)m_c^2}{\alpha(1-\alpha)}+\frac{48m_c^4}{\alpha^2M_B^2}\bigg\}\efun
\, .
\end{eqnarray}

For the interpolating current $J_{2\mu}$:
\begin{eqnarray}
\nonumber \rho^{pert}_2(s)&=&\frac{1}{2}\rho^{pert}_1(s),~~~~~
\rho^{\qq}_2(s)=\frac{1}{2}\rho^{\qq}_1(s),~~~~~
\rho^{\qq^2}_2(s)=\frac{1}{2}\rho^{\qq^2}_1(s) \, , \non
\rho^{\GGa}_2(s)&=&\frac{\GGb}{3^2\times2^{12}\pi^6}\dab \non
&&\bigg\{\frac{(1-\alpha-\beta)^2m_c^2}{\alpha^2}\bigg[\frac{48[(\alpha+\beta)m_c^2-2\alpha\beta
s]}{\alpha} -\frac{(5+\alpha+7\beta)\f(s)}{\beta^2}\bigg] \non
&&+\frac{6\f(s)}{\alpha\beta}\bigg[\frac{4(1-\alpha-\beta)(m_c^2-2\alpha\beta
s)}{\alpha} -m_c^2(1+\alpha+\beta)\bigg]\bigg\} \, , \non
\rho^{\qGq}_2(s)&=&-\frac{m_c^2m_q\qGq}{16\pi^4}\sqrt{1-4m_c^2/s}+\frac{\qGq}{3\times2^7\pi^4}\dab
\non
&&\bigg\{\frac{m_c(1-\alpha-\beta)[9(\alpha+\beta)m_c^2-19\alpha\beta
s]}{\beta^2}+\frac{12m_q
m_c^2}{\beta}-\frac{(11m_c-m_q)[3(\alpha+\beta)m_c^2-5\alpha\beta
s]}{\beta}\bigg\} \, ,
\end{eqnarray}
\begin{eqnarray}
\Pi^{\qGq\qq}_2(M_B^2)&=&\frac{\qGq\qq}{3^2\times2^5\pi^2}\int_0^1d\alpha\bigg\{3\alpha
M_B^2-\frac{2(6-7\alpha)m_c^2}{\alpha(1-\alpha)}+\frac{24m_c^4}{\alpha^2M_B^2}\bigg\}\efun
\, .
\end{eqnarray}

For the interpolating current $J_{5\mu}$:
\begin{eqnarray}
\nonumber
\rho^{pert}_5(s)&=&\frac{1}{3\times2^8\pi^6}\dab\frac{(1-\alpha-\beta)\f(s)^2}{\alpha^2\beta^3}
\bigg\{\frac{9(1+\alpha+\beta)\f(s)^2}{\alpha} \non
&&-\frac{4m_c^2(1-\alpha-\beta)(5+\alpha+\beta)\f(s)}{\alpha}
-108m_cm_q(1-\alpha-\beta)[(\alpha+\beta)m_c^2-3\alpha\beta
s]\bigg\} \, , \non \rho^{\qq}_5(s)&=&-\frac{\qq}{8\pi^4}\dab \non
&&\frac{\f(s)}{\alpha\beta}\bigg\{\frac{3m_c(1-\alpha-\beta)[3(\alpha+\beta)m_c^2-7\alpha\beta
s]}{\beta}-5m_q[(2-\alpha-\beta)m_c^2+\alpha\beta s]\bigg\} \, ,
\non
\rho^{\GGa}_5(s)&=&-\frac{\GGb}{3^2\times2^{12}\pi^6}\dab\bigg\{\frac{\f(s)m_c^2}{\alpha\beta}
\bigg[\frac{96(1-\alpha-\beta)^2(5+\alpha+\beta)}{\alpha^2} \non
&&+\frac{54(1-\alpha-\beta)(3+\alpha+\beta)}{\alpha}+\frac{5(1-\alpha-\beta)^2(29+13\alpha+13\beta)}{\alpha\beta}
+90(1+\alpha+\beta)\bigg] \non
&&+\frac{\f(s)s}{\alpha\beta}\bigg[288(1-\alpha-\beta)\alpha+60(1-\alpha-\beta)^2+120\alpha\beta\bigg]
\non
&&+\frac{32(1-\alpha-\beta)^2[(5-8\alpha-8\beta)m_c^2+18\alpha\beta
s]m_c^2}{\alpha^3}\bigg\} \, , \non
\rho^{\qGq}_5(s)&=&-\frac{m_q\qGq(s+26m_c^2)}{48\pi^4}\sqrt{1-4m_c^2/s}-\frac{\qGq}{3\times2^7\pi^4}\dab
\non
&&\bigg\{\frac{(1-\alpha-\beta)[87(\alpha+\beta)m_c^2-161\alpha\beta
s]m_c}{\beta^2}
+\frac{(14-71\alpha-14\beta)[3(\alpha+\beta)m_c^2-5\alpha\beta
s]m_c}{\alpha\beta} \non
&&-\frac{9m_q[(8-3\alpha-3\beta)m_c^2+5\alpha\beta
s]}{\beta}\bigg\} \, ,
\\
\rho^{\qq^2}_5(s)&=&\frac{\qq^2}{36\pi^2}\bigg[52m_c^2+2s-\frac{9m_qm_c(8m_c^2-3s)}{4m_c^2-s}\bigg]\sqrt{1-4m_c^2/s}
\, ,
\end{eqnarray}
\begin{eqnarray}
\nonumber
\Pi^{\qGq\qq}_5(M_B^2)&=&\frac{\qGq\qq}{96\pi^2}\int_0^1d\alpha\bigg\{(7\alpha+16)
M_B^2-\frac{2(8\alpha^2-25\alpha+4)m_c^2}{\alpha(1-\alpha)}+\frac{16(3-\alpha)m_c^4}{\alpha^2(1-\alpha)M_B^2}\bigg\}\efun
\, .
\end{eqnarray}

For the interpolating current $J_{6\mu}$:
\begin{eqnarray}
\nonumber \rho^{pert}_6(s)&=&\frac{1}{2}\rho^{pert}_5(s),~~~~~
\rho^{\qq}_6(s)=\frac{1}{2}\rho^{\qq}_5(s),~~~~~
\rho^{\qq^2}_6(s)=\frac{1}{2}\rho^{\qq^2}_5(s) \, , \non
\rho^{\GGa}_6(s)&=&-\frac{\GGb}{3^2\times2^{12}\pi^6}\dab\bigg\{\frac{\f(s)m_c^2}{\alpha\beta}
\bigg[\frac{48(1-\alpha-\beta)^2(5+\alpha+\beta)}{\alpha^2} \non
&&-\frac{18(1-\alpha-\beta)(3+\alpha+\beta)}{\alpha}+\frac{(1-\alpha-\beta)^2(29+13\alpha+13\beta)}{\alpha\beta}
+18(1+\alpha+\beta)\bigg] \non
&&+\frac{12[(1-\alpha-\beta)^2+2\alpha\beta]\f(s)s}{\alpha\beta}
+\frac{16(1-\alpha-\beta)^2[(5-8\alpha-8\beta)m_c^2+18\alpha\beta
s]m_c^2}{\alpha^3}\bigg\} \, ,
\\
\rho^{\qGq}_6(s)&=&-\frac{m_q\qGq(s+26m_c^2)}{96\pi^4}\sqrt{1-4m_c^2/s}-\frac{\qGq}{3\times2^7\pi^4}\dab
\non
&&\bigg\{\frac{(1-\alpha-\beta)[3(\alpha+\beta)m_c^2-13\alpha\beta
s]m_c}{\beta^2}-\frac{[(2+31\alpha-2\beta)m_c+3\alpha
m_q][3(\alpha+\beta)m_c^2-5\alpha\beta s]}{\alpha\beta}\bigg\} \,
, \non
\Pi^{\qGq\qq}_6&(M_B^2)&=\frac{\qGq\qq}{96\pi^2}\int_0^1d\alpha\bigg\{(8\alpha+11)
M_B^2-\frac{2(4\alpha^2-9\alpha-4)m_c^2}{\alpha(1-\alpha)}+\frac{8(3-\alpha)m_c^4}{\alpha^2(1-\alpha)M_B^2}\bigg\}\efun
\, .
\end{eqnarray}

From these results the expressions for the currents $J_{3\mu},
J_{4\mu}, J_{7\mu}$ and $J_{8\mu}$ can then be obtained conveniently
by the replacement $m_c\rightarrow -m_c$:
\begin{eqnarray}
\rho_1(s)\xrightarrow{m_c\rightarrow -m_c}\rho_3(s),
\rho_2(s)\xrightarrow{m_c\rightarrow -m_c}\rho_4(s),
\rho_5(s)\xrightarrow{m_c\rightarrow -m_c}\rho_7(s),
\rho_6(s)\xrightarrow{m_c\rightarrow -m_c}\rho_8(s) \, .
\end{eqnarray}
\end{document}